\documentclass[aip, jmp, amsmath, amssymb, reprint]{revtex4-1}

\usepackage{graphicx}
\usepackage{dcolumn}
\usepackage{bm}
\usepackage{color}
\usepackage[T1]{fontenc} 
\usepackage[english]{babel}
\usepackage{hyperref} % For hyperlinks in the PDF
\usepackage{siunitx} % SI units
\usepackage{fancyvrb}
\usepackage{booktabs} % Horizontal rules in tables
\usepackage{verbatim}
\usepackage{mathtools} % enlarges amsmath

\definecolor{ao(english)}{rgb}{0.0, 0.5, 0.0}
\newcommand{\black}[1]{\textcolor{black}{#1}}              % Valeria, December 2020
\begin{document}

\title{Ion acceleration by an ultrashort laser pulse interacting with a near-critical-density gas jet}

\author{M. Ehret}\email{michael.ehret@u-bordeaux.fr}
\affiliation{Universit\'{e} de Bordeaux, CNRS, CEA, CELIA (Centre Lasers Intenses et Applications), UMR 5107, Talence, France}
\affiliation{Institut f\"{u}r Kernphysik, Technische Universit\"{a}t Darmstadt, Darmstadt, Germany}

\author{C. Salgado-L\'{o}pez}\email{csalgado@clpu.es}
\affiliation{C.L.P.U. (Centro de L\'{a}seres Pulsados), Salamanca, Spain}
\affiliation{Universidad de Salamanca, Salamanca, Spain}

\author{V. Ospina-Boh\'{o}rquez}\email{valeria.ospina-bohorquez@u-bordeaux.fr}
\affiliation{Universit\'{e} de Bordeaux, CNRS, CEA, CELIA (Centre Lasers Intenses et Applications), UMR 5107, Talence, France}
\affiliation{Universidad de Salamanca, Salamanca, Spain}
\affiliation{CEA, DAM, DIF, F-91297 Arpajon, France}

\author{J.~A. Perez-Hern\'{a}ndez}
\affiliation{C.L.P.U. (Centro de L\'{a}seres Pulsados), Salamanca, Spain}

\author{M. Huault}
\affiliation{C.L.P.U. (Centro de L\'{a}seres Pulsados), Salamanca, Spain}
\affiliation{Universidad de Salamanca, Salamanca, Spain}

\author{M. de Marco}
\affiliation{C.L.P.U. (Centro de L\'{a}seres Pulsados), Salamanca, Spain}

\author{J.~I. Api\~{n}aniz}
\affiliation{C.L.P.U. (Centro de L\'{a}seres Pulsados), Salamanca, Spain}

\author{F. Hannachi}
\affiliation{CENBG, CNRS-IN2P3, Universit\'e de Bordeaux, Gradignan, France}

\author{D. De Luis}
\affiliation{C.L.P.U. (Centro de L\'{a}seres Pulsados), Salamanca, Spain}

\author{J. Hern\'{a}ndez Toro}
\affiliation{C.L.P.U. (Centro de L\'{a}seres Pulsados), Salamanca, Spain}

\author{D. Arana}
\affiliation{C.L.P.U. (Centro de L\'{a}seres Pulsados), Salamanca, Spain}

\author{C. Méndez}
\affiliation{C.L.P.U. (Centro de L\'{a}seres Pulsados), Salamanca, Spain}

\author{O. Varela}
\affiliation{C.L.P.U. (Centro de L\'{a}seres Pulsados), Salamanca, Spain}

\author{A. Debayle}
\affiliation{CEA, DAM, DIF, F-91297 Arpajon, France}
\affiliation{Universit\'e Paris-Saclay, CEA, LMCE, 91680 Bruy\`eres-le-Ch\^atel, France}

\author{L. Gremillet}
\affiliation{CEA, DAM, DIF, F-91297 Arpajon, France}
\affiliation{Universit\'e Paris-Saclay, CEA, LMCE, 91680 Bruy\`eres-le-Ch\^atel, France}

\author{T.-H. Nguyen-Bui}
\affiliation{Universit\'{e} de Bordeaux, CNRS, CEA, CELIA (Centre Lasers Intenses et Applications), UMR 5107, Talence, France}

\author{E. Olivier}
\affiliation{Universit\'{e} de Bordeaux, CNRS, CEA, CELIA (Centre Lasers Intenses et Applications), UMR 5107, Talence, France}

\author{G. Revet}
\affiliation{Universit\'{e} de Bordeaux, CNRS, CEA, CELIA (Centre Lasers Intenses et Applications), UMR 5107, Talence, France}

\author{N.~D. Bukharskii}
\affiliation{National Research Nuclear University MEPhI, Moscow, Russian Federation}

\author{H. Larreur}
\affiliation{Universit\'{e} de Bordeaux, CNRS, CEA, CELIA (Centre Lasers Intenses et Applications), UMR 5107, Talence, France}

\author{J. Caron}
\affiliation{Institut Bergoni\'{e}, D\'{e}partement de Radioth\'erapie, Unit\'{e} de Radiophysique, Bordeaux, France}

\author{C. Vlachos}
\affiliation{Universit\'{e} de Bordeaux, CNRS, CEA, CELIA (Centre Lasers Intenses et Applications), UMR 5107, Talence, France}
\affiliation{Centre for Plasma Physics and Lasers (CPPL), T.E.I. of Crete, Rethymnon, Greece}

\author{T. Ceccotti}
\affiliation{CEA/IRAMIS, SPAM, Gif-sur-Yvette, France}

\author{D. Raffestin}
\affiliation{Universit\'{e} de Bordeaux, CNRS, CEA, CELIA (Centre Lasers Intenses et Applications), UMR 5107, Talence, France}

\author{P. Nicolai}
\affiliation{Universit\'{e} de Bordeaux, CNRS, CEA, CELIA (Centre Lasers Intenses et Applications), UMR 5107, Talence, France}

\author{J.L. Feugeas}
\affiliation{Universit\'{e} de Bordeaux, CNRS, CEA, CELIA (Centre Lasers Intenses et Applications), UMR 5107, Talence, France}

\author{M. Roth}
\affiliation{Institut f\"{u}r Kernphysik, Technische Universit\"{a}t Darmstadt, Darmstadt, Germany}

\author{X. Vaisseau}
\affiliation{CEA, DAM, DIF, F-91297 Arpajon, France}

\author{G. Gatti}
\affiliation{C.L.P.U. (Centro de L\'{a}seres Pulsados), Salamanca, Spain}

\author{L. Volpe}
\affiliation{C.L.P.U. (Centro de L\'{a}seres Pulsados), Salamanca, Spain}
\affiliation{Laser-Plasma Chair at Universidad de Salamanca, Salamanca, Spain}

\author{J.~J. Santos}\email{joao.santos@u-bordeaux.fr}
\affiliation{Universit\'{e} de Bordeaux, CNRS, CEA, CELIA (Centre Lasers Intenses et Applications), UMR 5107, Talence, France}

\begin{abstract}

We demonstrate laser-driven Helium ion acceleration with cut-off energies above \SI{25}{\mega\electronvolt} and peaked ion number above \SI{e8}{\per\mega\electronvolt} for \SI{22+-2}{\mega\electronvolt} projectiles from near-critical density gas jet targets. We employed shock gas jet nozzles at the high-repetition-rate (HRR) VEGA-2 laser system with \SI{3}{\joule} in pulses of \SI{30}{\femto\second} focused down to intensities in the range between $9\times 10^{19}\,\rm Wcm^{-2}$ and $1.2\times 10^{20}\,\rm Wcm^{-2}$. We demonstrate acceleration spectra with minor shot-to-shot changes for small variations in the target gas density profile. Difference in gas profiles arise due to nozzles being exposed to a experimental environment, partially ablating and melting.

\end{abstract}

\maketitle

% -----------------------------------------------------------------------------------------------------------------------------------------------------------------------------
\section{Introduction}

Since the advent of high-power, short-pulse lasers \cite{Strickland_1985}, much effort has been devoted to the development of high-energy, high-quality laser-driven ion sources \cite{Daido_2012,Macchi_2013}. The interest in these sources stems from their many potential applications, including time-resolved probing of laser-induced phenomena \cite{Borghesi_2004, Santos_2015, Eh2017-1}, isochoric heating of dense plasmas \cite{Patel_2003}, fast ignition of inertial confinement fusion targets \cite{Roth_2001}, nuclear physics \cite{Ledingham_2003} or medical purposes \cite{Spencer_2001, Lendingham_2014}. Most of these applications exploit the unique properties of laser-driven ion beams, notably their short duration, high number density, high laminarity
and compactness.

The laser interactions giving rise to ion acceleration can occur in two main regimes, depending on the transparency/opacity properties of the irradiated (and ionized) medium. 
Solid or liquid targets with electron densities ($n_e$) larger than the critical density ($n_c = 1.7 \times 10^{21}\,\rm cm^{-3}$ for the central $0.8\,\mu \rm m$ wavelength of a Ti:Sa laser)
are opaque to the laser light, yet a significant ($\sim 1-50\,\%$) fraction of the laser energy can convert into high-energy ($\sim \rm MeV$) electrons at the target surface. When propagating
across the target, these fast electrons can induce strong charge-separation fields at the target/vacuum interfaces capable of accelerating the surface ions (mainly protons) to high
(up to $\sim 100\,\rm MeV$) energies \cite{Wagner_2016, Higginson_2018}. This process, referred to as target normal sheath acceleration (TNSA), is the most widely studied and exploited
method for laser-based ion acceleration \cite{Daido_2012, Macchi_2013}.

In dilute gaseous targets ($n_e \ll n_c$), the laser pulse can propagate large distances through the plasma while generating a nonlinear wakefield that is able to accelerate
some bulk electrons to ultrarelativistic energies \cite{Esarey_2009}. Simultaneously, the transverse ponderomotive force of the focused laser pulse tends to expel radially
the bulk electrons, thus creating a positively charged plasma channel. The related electrostatic fields can accelerate the plasma ions to MeV-range energies in the transverse plane
\cite{Krushelnick_1999, Willingale_2006, Lifschitz_2014}.

Fewer studies have addressed the case of short laser pulses interacting with plasmas of electron density $n_e \sim 0.1-1\,n_c$, mostly due to the experimental difficulty of
achieving such conditions in a controlled manner. This so-called near-critical interaction regime, however, is spurring increasing interest because it is predicted to optimize the energy
coupling between the laser and the plasma \cite{Willingale_2009, Fiuza_2012, Debayle_2017, Rosmej_2019, Pazzaglia_2020}. An intense laser pulse impinging onto a near-critical
plasma can drive a strong charge separation in the longitudinal direction, launching an electrostatic shock wave \cite{Moiseev_1963, Forslund_1970, Sorasio_2006, Cairns_2014}. 
The supersonic propagation of this structure through the plasma goes along with partial reflection of the background ions at twice the shock velocity, a mechanism termed collisionless shock
acceleration (CSA). The charge separation underpinning the shock formation can be driven either directly by the laser, via its ponderomotive push on the opaque region (if any) of the
plasma profile \cite{Silva_2004, Fiuza_2012, Fiuza_2013, Kim_2016, Liu_2016, Pak_2018}, or indirectly, via the pressure gradients associated with the laser heating of the bulk electrons
in a fully transparent, nonuniform plasma \cite{Debayle_2017, Antici_2017, Chen_2017,Puyuelo_2019}. Depending on the gas profile, CSA may come along with additional acceleration mechanisms,
such as TNSA \cite{Fiuza_2012,Fiuza_2013} or magnetic vortex acceleration \cite{Bulanov_2010, Nakamura_2010, Sylla_2012_2, Helle_2016, Park_2019}.

%This acceleration mechanism, named as Collisionless Shock Acceleration (CSA), is predicted to generate narrow and spectrally peaked ion beams, up to 100s MeV of energy.

The laboratory production of near-critical plasmas remains a nontrivial task, especially when high-repetition-rate ion sources are needed. The use of widely available near-infrared ultrashort
pulses requires electron plasma densities of the order of $10^{20-21}\,\rm cm^{-3}$, which nowadays can be achieved by means of exploding foils (using either the pedestal of the
laser pulse or an additional pulse)~\cite{dHumieres_2013_1, Gauthier_2014, Antici_2017, Pak_2018}, foam targets~\cite{Willingale_2009, Prencipe_2016, Rosmej_2019},
nanowire arrays \cite{Bin_2015, Bin_2018}, or high-density gas jets~\cite{Goers_2015, Chen_2017, Henares_2019}. In this paper, we report on experimental measurements of the
interaction of an ultrashort (30~fs), ultraintense ($\sim 10^{20}\,\rm Wcm^{-2}$) laser pulse with a high-density gas jet which, once fully ionized, can attain a maximum electron
density close to the critical density \cite{Sylla_2012_1}.

%Where high power laser pulses from $\mathrm{CO_2}$ lasers with wavelength \SI{10.6}{\micro\metre} correspond to a critical density of $n_c \approx 10^{19} \mathrm{cm^{-3}}$, that may be reached with a standard gaseous target \cite{Haberberger_2012},} the development of the new generation of \black{HRR} high energy near infra-red (NIR) laser systems \black{with wavelength in the \si{\micro\metre}-range demands higher target densities. This} has lead to an effort on \black{novel} targetry strategies\black{.} Acceleration of ions driven by short pulse NIR lasers have been demonstrated by using exploding foils \cite{dHumieres_2013_1, Gauthier_2014, Antici_2017, Pak_2018}, foam targets \cite{Willingale_2009, Prencipe_2016, Rosmej_2019}, or high-density gas jets \cite{Goers_2015, Chen_2017, Henares_2019}. This kind of targets have been proposed as well as targets for electron beam acceleration \cite{Goers_2015,Rosmej_2019} and positron \cite{Liu_2018} and gamma-ray \cite{Brady_2013} generation. 

%The presence of several simultaneous ion acceleration mechanisms, the non-trivial handling of these special targets, and the presence of many different non-linear plasma effect\black{s}, such as relativistic self-focusing \cite{Sylla_2013}, magnetic vortex generation \cite{Sylla_2012_2,Nakamura_2010} and magnetic self-channelling \cite{Pukhov_1996}, make the experiments specially complex in terms of diagnostics dimensioning, \black{allowing one to discriminate between all mechanisms at play}.

This paper is organized as follows. The experimental setup is explained in Section \ref{Experimental setup}. The main experimental results, divided in two different studies, are presented in detail in Section \ref{Experimental results}. The particle-in-cell (PIC) simulation results are summarized in Section \ref{PIC Simulations}. Finally, the conclusions and foreseen applications are outlined in Sections \ref{Conclusion} and \ref{Applications}, respectively. Additional material regarding laser diagnostics and data analysis can be found in Section \ref{sec:methods}.

% -----------------------------------------------------------------------------------------------------------------------------------------------------------------------------

\section{Experimental setup}
\label{Experimental setup}

\begin{figure*}[htb]
\centering
\includegraphics[width=0.95\linewidth]{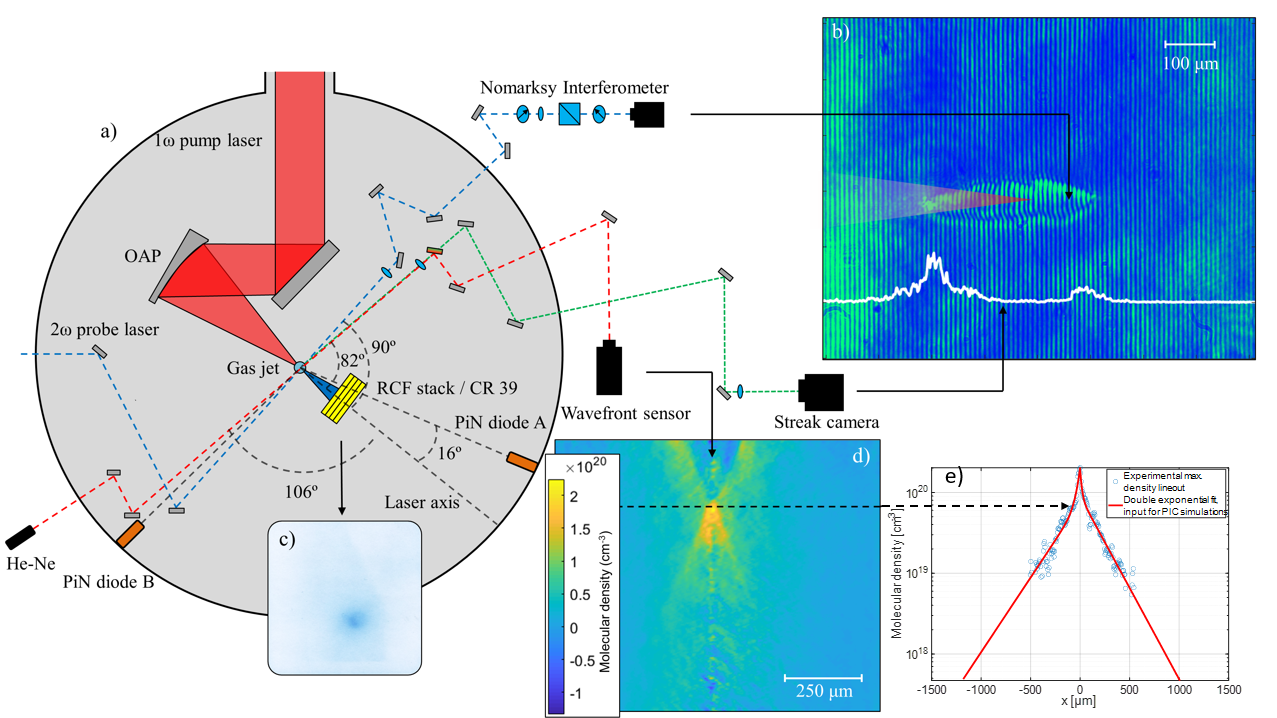}
\caption{(a) Experimental setup top view. (b) Raw data of the interferometer. Overlaid are a lineout of the plasma self-emission as measured (arb. units) by the streak camera (white curve)
and a sketch of the laser envelope in vacuum (transparent red triangle). The laser focal spot in vacuum coincides with the shocked gas region. (c) Typical RCF signal. (d) Molecular density map of the 9/1 Nitrogen/Helium molecular density ratio gas jet prior to interaction, as measured by the wavefront sensor (nozzle output is located 300 $\mu m$ above the image). (e) Molecular density lineout at the shock height (circles) and double exponential fit of the measured data (solid line) extrapolated until $n_{mol,min} = 5\times 10^{17}\,$cm$^{-3}$ for numerical purposes (see Section \ref{PIC Simulations}).}
\label{fig:setup}
\end{figure*}

%TODO: shot-to-shot fluctuations of laser energy as plus-minus

\begin{figure}[htb]
\centering
\includegraphics[width=\columnwidth]{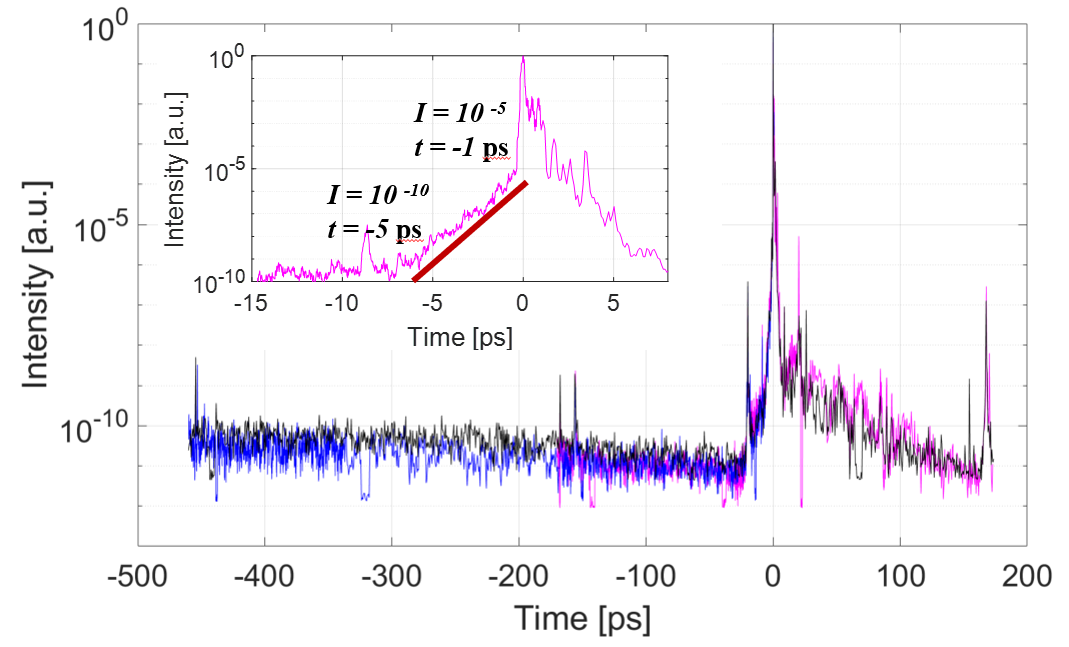}
\caption{\black{Three independent third-order autocorrelator (Sequoia) measurements of the temporal contrast at best compression with a \numrange{e-10}{e-11} ASE level. Inset zooms in the $\approx$ 5 ps long intensity ramp prior to the arrival of the main intensity peak. For more details see Section \ref{Additional Material: Laser Diagnostics}}}
\label{fig:contrast_with_inset}
\end{figure}

% Facility
The experiment was performed using the VEGA 2 Ti:Sa laser system at the Centro de L\'aseres Pulsados (CLPU) \cite{Volpe_2019} in September 2018. The laser pulse, of 3 $\pm$ 0.36 J energy, 30~fs duration and 
$0.8\,\mu m$ wavelength, was focused by a $F/4$ off-axis parabolic (OAP) gold-coated mirror into the gas, see Fig.~\ref{fig:setup}(a). The focal spot was consistently measured
at low energy by a high magnification imaging system. A reproducible full width at half maximum (FWHM) of $7\,\rm \mu m$ was obtained, yielding a maximum intensity between $9\times 10^{19}\,\rm Wcm^{-2}$ and $1.2\times 10^{20}\,\rm Wcm^{-2}$ at full pulse energy. The amplified stimulated emission (ASE) was characterized by an intensity contrast of $\num{\sim 5e-12}$ at $t=-100\,\rm ps$
(before the pulse maximum), as measured through third-order autocorrelation, see Fig.~\ref{fig:contrast_with_inset}. The contrast was found to be $\sim 8\times 10^{-9}$ at $t=-10\,\rm ps$ and $\sim 5\times 10^{-8}$ at $t=-5\,\rm ps$, prior to the high intensity peak.

We employed a high-pressure gas jet target developed by SourceLab \cite{Sylla_2012_1}, comprising a shock-geometry nozzle attached to a rapid electro-valve
with 400~bar of gas backing pressure. This yielded a narrow density profile of approximately $100\,\rm \mu m$ FWHM across the shocked region at a height of $\approx$ 470 $\mu m$ in respect to the nozzle extremity surface [see Figs.~\ref{fig:setup}(d) and (e)]. 
At full ionization, the peak electron density reached $n_e= 2.8 \times 10^{21}$ cm$^{-3}$, that is, slightly above the critical density associated with a $0.8\,\rm \mu m$ laser. Before each shot, the gas jet was
characterized through interferometry using the commercially available SID-4 wavefront sensor (PHASICS) \cite{Chanteloup_2005}. The nozzle was also visually inspected so as to ensure that
neither its surface nor its cylindrical geometry had been altered. 
%The shot-to-shot characterization \black{to} evaluate \black{the} laser damage \black{of the nozzle and its effect on} the gas jet profile.
Several types of gas were studied: Helium (He), Nitrogen (N), and a 9/1 Nitrogen/Helium molecular density ratio gas mixture. The addition of N, due to its fluid properties, proved to be beneficial for
the proper operation of the valve.
%The Helium doping \black{yields ions with high charge-to-mass ratio} in the bulk of the plasma that \black{are expected} more susceptible to acceleration (only true if N atoms are partially ionized).
A three-axis motorization system for the valve holder and two perpendicular imaging systems allowed the density peak to coincide with the laser spot in vacuum.

% Diagnostics
\black{Main diagnostics are on-shot interferometry to determine the driven plasma density, the streaked imaging of plasma self-emission to identify hot plasma regions, and passive particle detectors as well as HRR ready time-of-flight (ToF) detectors to measure ion beam spectra.}

%Pick-up probe intro
\black{Frequency-doubled ultra-short pick-up probe is derived from the edge of the main laser pulse. Synchronization with the main beam is done with the help of a delay line and a streak camera. We observe the plasma dynamics} perpendicularly to the main \black{driver} beam axis, around \SI{1}{\nano\second} after its arrival. \black{One half of the probe beam is aligned to propagate through free space, the other portion encounters a different optical path going through the density profile. The resulting phase shift} was measured with a Nomarsky interferometer\black{, see figure \ref{fig:setup} b)}. Knowing the relation between index of refraction of a plasma and its density, and imposing cylindrical symmetry around the laser axis, \black{one deconvolutes} the ultrafast time-resolved \black{density map} of the plasma \cite{Hutchinson_2002} [see Fig.~\ref{fig:setup}(b)].

%Streak camera intro
A fast streak camera \black{(Hamamatsu C7700), with \si{\pico\second} resolution,} measured the plasma self-emission \black{under} an angle of \SI{82}{\degree} with respect to the \black{main laser} axis \black{in the horizontal plane, blocking the NIR photons with a BG38 filter. This restriction} to optical photons \black{is a measure to protect the streak camera from} scattered \black{light of the driver laser}. \black{A line-out of the \black{prompt} self emission in arbitrary units is superposed to the interferometry in figure \ref{fig:setup} b). Noteworthily, self emission is prominent at both ends of the plasma channel, but not at the position where the laser focus was encountered without gas.}

%TOF intro
\black{O}n-line energy spectra of the accelerated particles \black{were acquired by two Silicon PIN photodiode ToF detectors.} \black{Both photodiodes were} located \black{vertically inclined with \SI{19}{\degree}} at \SI{67}{\centi\metre} from \black{the interaction point at different horizontal angles of (A)} \SI{16}{\degree} and \black{(B)} \SI{106}{\degree} \black{with respect to the driver laser propagation axis (see figure \ref{fig:setup}.a.} \black{The diode substrate is preceded by a Silicon layer of \SI{750}{\nano\metre} and extra filtering by Mylar foils with \SI{40}{\nano\metre} thick Aluminium coating,} in order to avoid diode saturation \black{by the} strong photopeak \black{from the} laser-target interaction. \black{Mylar foils were (A)} \SI{2}{\micro\metre} \black{and (B)} \SI{4}{\micro\metre} \black{thick}. \black{S}ignal acquisition \black{with a \SI{1}{\giga\hertz} oscilloscope \black{(Tektronix DPO4104)} was performed at} a maximum \black{sampling rate of \SI{200}{\pico\second} per sample.} ${20\mathrm{~dB}}$ attenuation \black{was applied to} both channels. \black{For details on data analysis, see additional material in section \ref{sec:methods}--\ref{sec:material_ToF}.}

%Passive particle detector intro
\black{We employed solid state passive particle detectors} to capture the spatial and spectral properties of \black{forward accelerated} ion beams \black{in single-shots}, with stacks of Radiochromic Films (RCF) \cite{Nue2009} \black{and slaps of Columbia Resin \#39 (CR-39) \cite{Yo1958,DB1987}. RCF undergo a color-changing radio-synthesis that allows to retrieve the dose-depth curve and therewith an absolute projectile number spectrum, given the projectile species is known. To identify the projectile species, we make use of CR-39 as solid state track detector. CR-39 is especially used to distinguish ions from electrons, as projectiles with a low linear energy transfer (LET) of \SIrange{10}{15}{\kilo\electronvolt\per\micro\metre} from projectile to detector, typical for single electrons, do not lead to the formation of tracks \cite{Ha2008}.}

\black{As RCF} we used two different types of GAFCHROMIC  film\black{(Ashland)}, commercially available EBT-3 \cite{GEBT3} and specially manufactured U-EBT-3\footnote{Films were made available to us by the Advanced Materials Group of Ashland Specialty Ingredients G.P., 1005 US Hwy No 202/206, Bridgewater, NJ 08807, USA.}. \black{For details on RCF data analysis, see section \ref{sec:methods}--\ref{sec:material_RCF}. Each RCF stack comprises $4$ layers of U-EBT-3 and $2$ layers of EBT-3 and is enveloped in \SI{10}{\micro\metre} Aluminium filter foil. Active layers of U-EBT-3 face the interaction point.} \black{CR-39 were of type TasTrak CR-39 Plastic Sheet bought in February 2015. CR-39 are enveloped in} an opaque Aluminium coated \SI{2}{\micro\metre} thick Mylar foil. \black{For more details on CR-39 data analysis, see \ref{sec:methods}--\ref{sec:material_CR39}.} 

The passive particle detector surfaces are aligned \SI{60}{\milli\metre} from \black{the interaction point} - and detector slaps are perpendicular to the laser axis.

\begin{figure}[htb]
\centering
\includegraphics[width=\columnwidth]{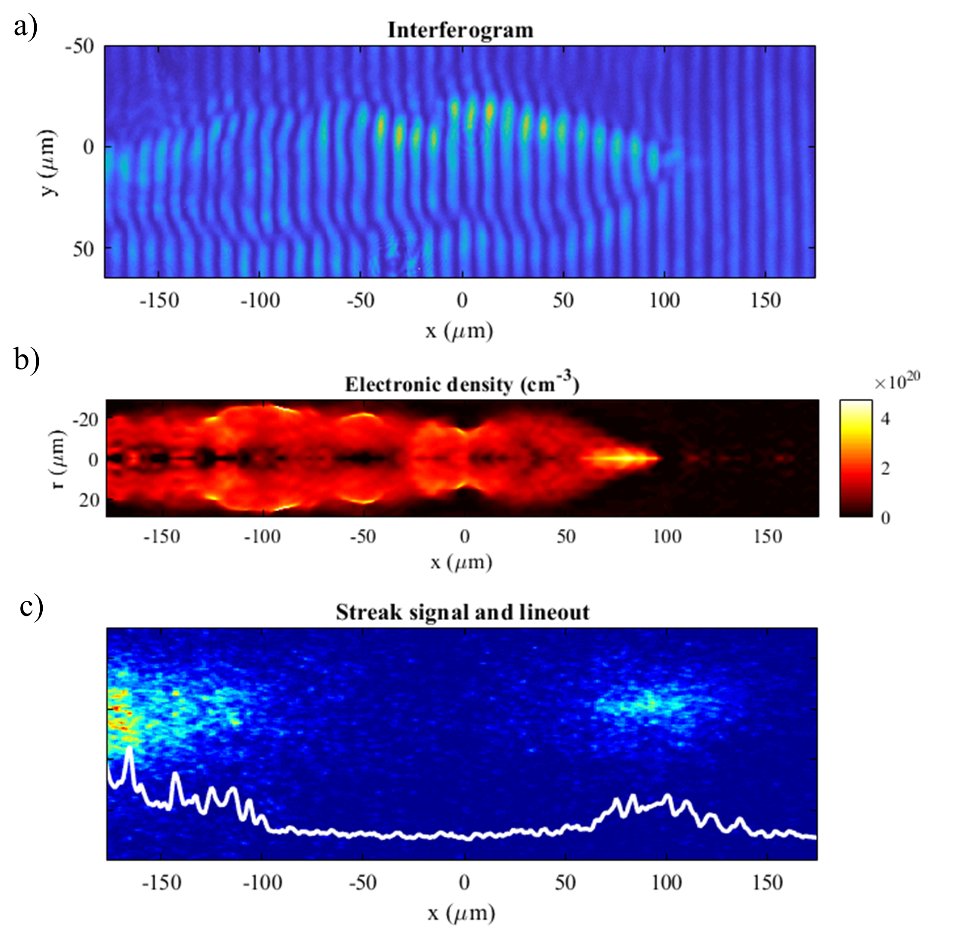}
\caption{Optical plasma diagnostics show strong self emission from dense narrow regions of the plasma. We see (a) raw interferogram measurement of the plasma, 1 ns after the interaction starts, measured with the Nomarsky interferometer (b) the retrieved electronic density of the plasma and (c) the raw streak camera measurement of the self emission of the same laser shot (including measurement lineout in white). $x=0,y=0$ are the coordinates of the jet shock point; the laser travels left to right and focalizes in the same point.}
\label{optical_diag}
\end{figure}

% -----------------------------------------------------------------------------------------------------------------------------------------------------------------------------

\section{Experimental results}
\label{Experimental results}

\black{For laser shot \#63, deploying a pure helium jet}, figure \ref{optical_diag} illustrates results of optical diagnostics, with \ref{optical_diag} (a) the raw interferometric data, \ref{optical_diag} (b) the electronic density map and \ref{optical_diag} (c) the streaked self-emission of the plasma created \black{in the same laser shot}. As expected, taking into account that the interferometry measurement is made \SI{1}{\nano\second} after the interaction, the plasma has been able to expand and the average density is a fraction of the critical density\black{, $\approx 0.1 n_c$}. \black{The plasma is fully ionized, the retrieved electron number density at the end of the plasma channel is twice as high as the un-driven gas density at the same position for shot \#63.} \black{The plasma self-emission signal appears frozen in \ref{optical_diag} (a), meaning that its duration is much shorter than the resolution of the streak with a \SI{2}{\nano\second} streak window.}

\begin{figure*}[htb]
\centering
\includegraphics[width=\linewidth]{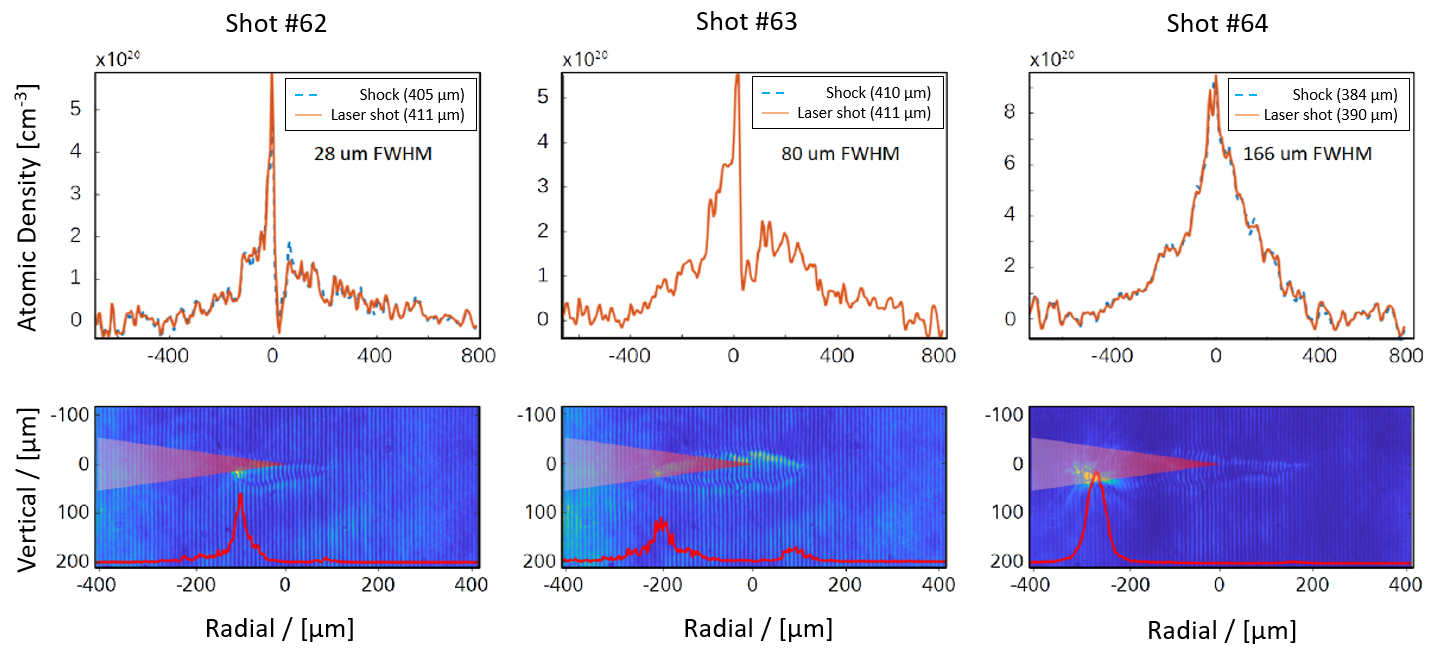}
\caption{\black{For three consecutive shots varying the pure Helium gas jet density profile, (top row) longitudinal un-driven gas jet density profile, (bottom row) interferometric image superposed to aligned laser beam and plasma self-emission in arbitrary units as red line. The laser is focused to $x=0,y=0$, the coordinates of the jet maximum density point.}}
\label{gas_62_63_64_overview}
\end{figure*}

\black{In figure \ref{gas_62_63_64_overview} (bottom row) the} lineout of the signal of the self-emission of the plasma is overlaid to a picture of the raw data of the interferometer. The scaled line-out of the self-emission, \black{with position indicated in \ref{optical_diag} (a),} \black{is visible as white line in \ref{optical_diag} (c)}. This data clearly shows that the laser goes through the peak of the density profile and deposits energy downstream.

\subparagraph{First study}

Laser shots \#62, \#63 and \#64 correspond to different gas jet density profiles, as shown in figure \ref{gas_62_63_64_overview} (top row), \black{obtained for different gate opening times of the gas reservoir. Note that the backing pressure of \SI{400}{\bar} was constant throughout the experiment}. One observes a plateau like feature in the density profile on the left hand side, right before the density peak, see Fig. \ref{gas_62_63_64_overview} (\#62, top row). The position of the plateau and the location of the self emission peak coincide for shot \#62. The plateau density rises \black{in} shot \#63 and evolves to \black{become part of the rising edge of} the peak for shot \#64. Thus, for this series of shots, the laser has to traverse a rising areal density of gas before reaching the peak. There, a second self emission zone appears in vicinity of the end of the plasma channel. Comparing Fig. \ref{gas_62_63_64_overview} (top row) and Fig. \ref{gas_62_63_64_overview} (bottom row), we observe that the first self emission peak corresponds to a initial atomic density of \SIrange{1.2e20}{1.6e20}{\per\cubic\centi\metre}.

\begin{figure*}[htb]
\centering
\includegraphics[width=\linewidth]{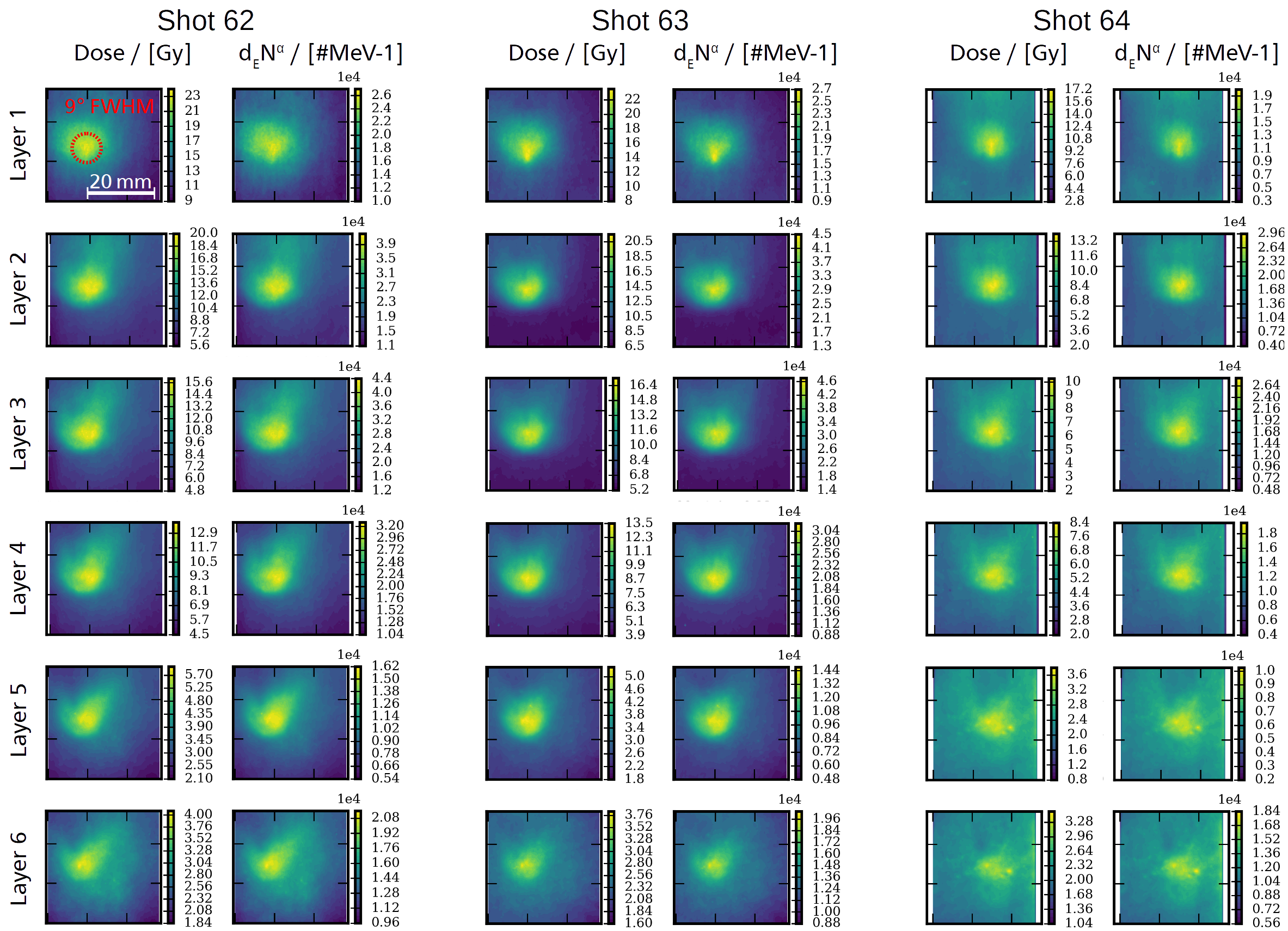}
\caption{\black{RCF imprints converted to dose and alpha particle number density for three consecutive shots varying the gas jet density profile. The number density conversion presumes the appearance of alpha particles, a cut-off energy equal to the last signal layer and a step-wise flat spectral shape between layers. The typical FWHM divergence angle of the beam imprint is \SI{9}{\degree}, all images have the same spatial scale. Note that number densities are normalized to one pixel, with respect to the resolution of RCF scans of $600\mathrm{~dpi}$. For shots \#62 and \#63, the presumed pre-aligned laser axis corresponds to the centre of the illustrated frames. For shot \#64 this does not apply.}}
\label{RCF_62_63_64_data}
\end{figure*}

\black{Dose converted RCF data is \black{shown} in figure \ref{RCF_62_63_64_data} for these 3 shots, \black{laser photons and particles propagate out of the sketch plane}. One sees active layer imprints over the full RCF stack, layers are numbered in ascending order in direction of particle propagation. A beam-like feature with similar divergence and amplitude is clearly pronounced in all shots. The average FWHM divergence of the beam is \SI{9}{\degree}. The presumed laser axis according to pre-alignment coincides with the centre of depicted frames for shots \#62 and \#63, its position is not superposed with the maximum measured dose deposition. This may be due to a alignment error, the RCF stack is not thick enough to confidently determine the particle beam trajectory.}

\black{The deposited dose decreases \black{slowly} throughout the stack. \black{This does not resemble a dose-depth curve expected for high energy electrons \cite{Me1996}, which would have a peak dose within the stack followed by a steep decay.} We derive ion number densities per pixel following a standard protocol established in \cite{Nue2009}, presuming alpha particle projectiles, a beam cut-off directly after the last layer of the stack and a step-wise flat spectral shape between two active layers. Note that the number densities in figure \ref{RCF_62_63_64_data} are relative to one pixel, with scanner resolution of $600\mathrm{~dpi}$.}

\begin{figure}[htb]
\centering
\includegraphics[width=\columnwidth]{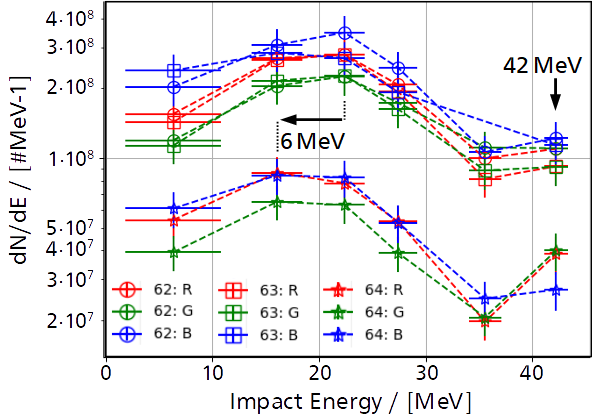}
\caption{\black{Projectile spectrum for three consecutive shots varying the gas jet density profile, presuming an alpha particle beam impacting on the stack of RCF.  RCF color channels R, G and B are independently analyzed.}}
\label{RCF_62_63_64_spectrum}
\end{figure}

\black{With no further \black{diagnostic of} the background signal issued by high energy electrons and photons, we presume the signal of an alpha particle beam with small angular divergence superposed to a broad background. Number density maps are fitted with the sum of two 2D-Gaussian functions for the particle beam, superposed with a second order polynomial in both x- and y-dimension representing the background signal. Fits do not converge without a weighting of the data: after visual centering of the peak signal on a fit-canvas, the uncertainty of pixels is linearly and concentrically increased from $100 \text{\%}$ at the peak to $167 \text{\%}$ in the corners. No further adjustment is done. The integrated particle number within both \black{Gaussians} is \black{plotted} in figure \ref{RCF_62_63_64_spectrum}.}

\black{The alpha particle spectrum has a peaked shape with following exponential decay. Note that the increase at the end of the spectrum is artificial due to the assumption of a cut-off energy of \SI{42}{\mega\electronvolt}, yielding an overestimation of the number density influenced by all higher energy projectiles that may have been present. The spectra for shots \#62 and \#63 are similar in the range of their error. The number density maximum of \SIrange{2e8}{3e8}{\per\mega\electronvolt} is attained for \SI{22+-2}{\mega\electronvolt} alpha particles. Note that both shots have comparable neutral gas density peaks. Shot \#64, with higher gas density peak, shows a thoroughly lower particle number density and the spectral peak at lower energy. \SI{8e7}{\per\mega\electronvolt} are attained for \SI{16+-2}{\mega\electronvolt} alpha particles.}

\begin{figure}%[ht]
\centering
\includegraphics[width=\columnwidth]{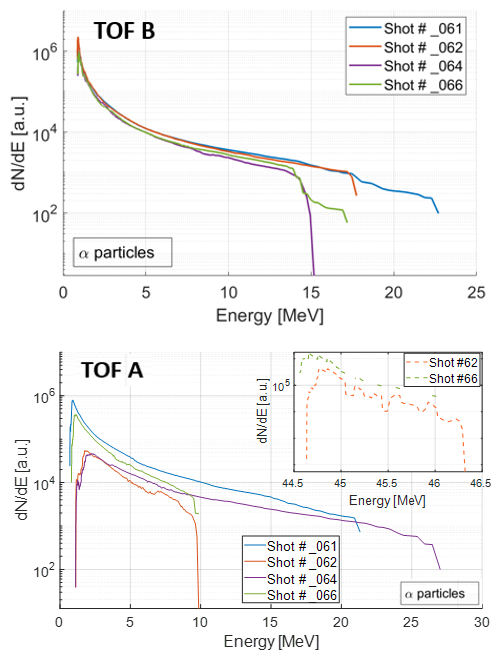}
\caption{\black{ToF results presuming alpha particle acceleration yield similar spectra in transverse direction (photodiode B, top panel) when comparing shots with similar plasma and self emission characteristics, here for pairs \#61 -- \#62 and \#64 -- \#66. With results obtained under a small angle in laser-forward direction (photodiode A, bottom panel) we see the influence of the passive particle detector present for shots \#62 and \#64 that acts as a thick filter with low energy cut-off of \SI{44}{\mega\electronvolt}.} \black{Ions stopping over the "filter" shift the maximum detected energy down to $\approx$ 10 MeV in the case of shots \#62 and \#66. Inset shows spectra of shots \#62 and \#66 considering the attenuation in the RCF stack, see details of the stack composition in Section \ref{Experimental setup}.}}
\label{spectra_alpha_61_62_64_66}
\end{figure}

\black{In order to better work out the difference in the origin of the different spectra, we turn to the ToF detector results.} The deconvolved spectra obtained for \black{the transversely oriented photodiode B} assuming alpha \black{projectiles} is plotted in figure \ref{spectra_alpha_61_62_64_66} (top panel).

\black{Note that the longitudinal profile for the un-driven gas jet, the interferometric image of the driven plasma and the plasma self emission are similar for the pairs of shots \#61 -- \#62 and \#64 -- \#66. For smaller volumes of plasma and lower self-emission amplitude in \#61 -- \#62, the particle number and high energy cut-off of the transversely accelerated particles are higher than for larger volumes of plasma and higher self-emission amplitude in \#64 -- \#66. The decrease of particle number and the shift of spectral features to lower energies, here detected in a direction practically perpendicular to the intense laser axis, aligns with the forward accelerated spectra retrieved with RCF.}

\black{The ToF spectra for alpha particles accelerated under a small angle from the laser forward direction are shown in Fig. \ref{spectra_alpha_61_62_64_66} (bottom panel). The deployed RCF stack in shots \#62 and \#64 acts as a high-pass filter and cuts off energies below \SI{44}{\mega\electronvolt}. Overall, alignment precision of the recesses holding the RCF was not well enough monitored to exclude possible clipping of the direct line between laser-focus and ToF photodiode A. Results have to be understood as a lower limit to the unperturbed projectile energy. Therefore, the below limit for the highest recorded projectile energy calculates to \SI{55}{\mega\electronvolt} for alpha particles.}

\subparagraph{Second study}

%Next Insights: RCF
\black{A second study aims at the influence of the transverse laser focus position with respect to the gas jet density maximum. \black{We} deploy \black{a gas} mixture of Nitrogen and Helium in a $9$ to $1$ molecular density ratio. The gas jet density was not deliberately altered, but changed with the progressive destruction of the nozzle on a shot-to-shot basis.}

\begin{figure*}[htb]
\centering
\includegraphics[width=\linewidth]{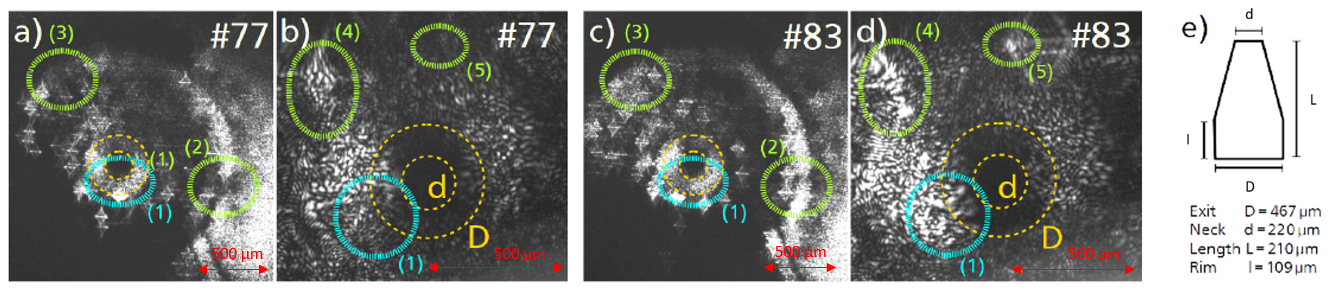}
\caption{\black{The shock nozzle deployed for the series of shots \#77 to \#87 shows visual changes  after some laser shots. The bottom view image of the original nozzle (a) and a magnified image (b) before shot \#77  are compared to the respective images taken before shot \#83 in (c) and (d). The shock nozzle geometry (e) is initially cylindrical symmetric and mechanical changes to the shape and the surface quality impact the gas flow. Note visual changes in (a--d) with highlighted areas (1--5), of which (1), in blue, affects directly the nozzle cone (exit and neck diameter are indicated with yellow dashed lines).}}
\label{nozzle_77_83_TUCAN}
\end{figure*}

\begin{figure*}[htb]
\centering
\includegraphics[width=\linewidth]{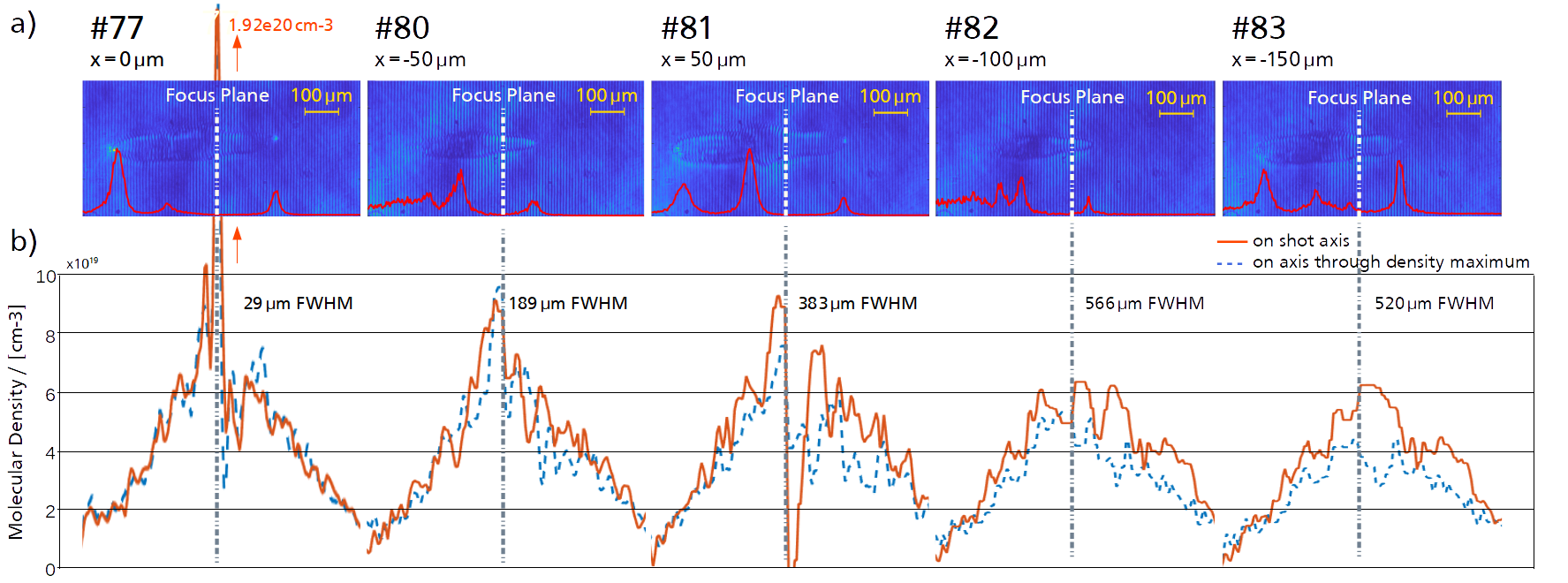}
\caption{\black{(a) Interferometric images overlaid by self-emission line-outs for shots with variation of the transverse position of the laser-gas interaction. The longest plasma channel is recorded for shot \#81 with positive transverse displacement -- the length of the channel decreases for the series of shots towards negative transverse displacements. The last shot of the series \#83 does not agree with this assessment and shows a longer channel again. Note that (b) un-driven gas density profiles acquired prior to corresponding shots show visual changes from shot-to-shot, indicating changes to the nozzle. This agrees with a qualitative difference of the nozzle surface between shot \#77 and \#83, presented with fig. \ref{nozzle_77_83_TUCAN}.}}
\label{gas_77_80_81_82_83}
\end{figure*}

\begin{figure*}[htb]
\centering
\includegraphics[width=\linewidth]{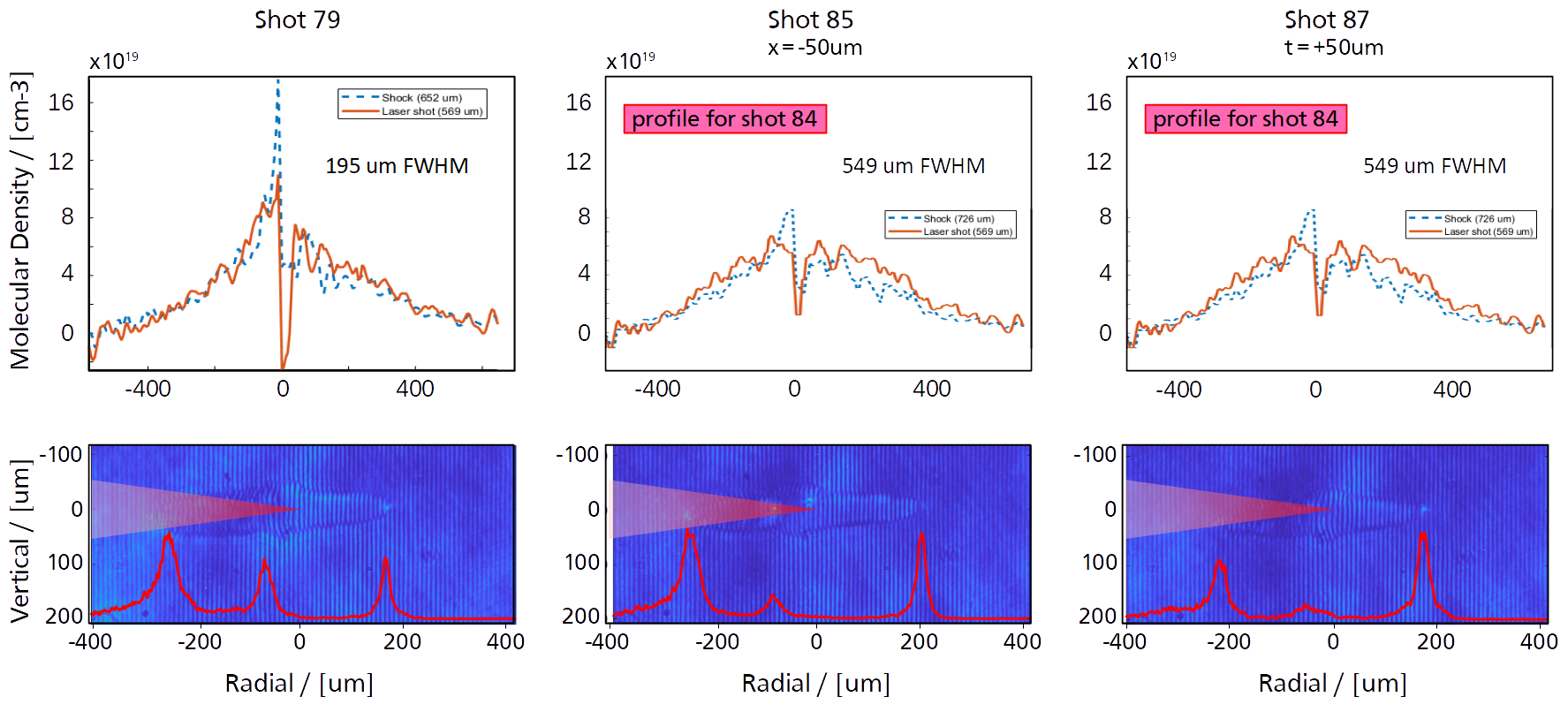}
\caption{\black{Shot-to-shot variation of the gas jet density profile using a mix of Helium and Nitrogen in a ratio $1$ to $9$, (top row) longitudinal un-driven gas jet density profile, and (bottom row) interferometric image superposed to aligned laser beam and plasma self-emission in arbitrary units as red line, \SI{1}{ns} after the interaction starts. The laser is focused to $x=0,y=0$, the coordinates of the jet shock point, but transverse displaced by \SIlist{0;\pm50}{\micro\metre}.}}
\label{gas_79_85_87_overview}
\end{figure*}

\black{Damages were noticed visually with the nozzle viewing system, see figure \ref{nozzle_77_83_TUCAN}, and due to alteration of the un-driven gas jet density profile. From gas profiles in figures \ref{gas_79_85_87_overview} (top row) and \ref{gas_77_80_81_82_83} (b), we infer likely nozzle damage by laser shots \#79 and \#81. Interferometric data in figure \ref{gas_77_80_81_82_83} (a) shows a particularly long plasma channel even though the laser was focused far from the largest areal density in vicinity of the supposed gas density peak. This hints severe damage to the nozzle that destroyed the cylindrical symmetry of the gas jet with shot \#82. Areal density and peak density of the gas target are thoroughly lower than in the previous study.}

\begin{figure*}
\centering\includegraphics[width=\textwidth]{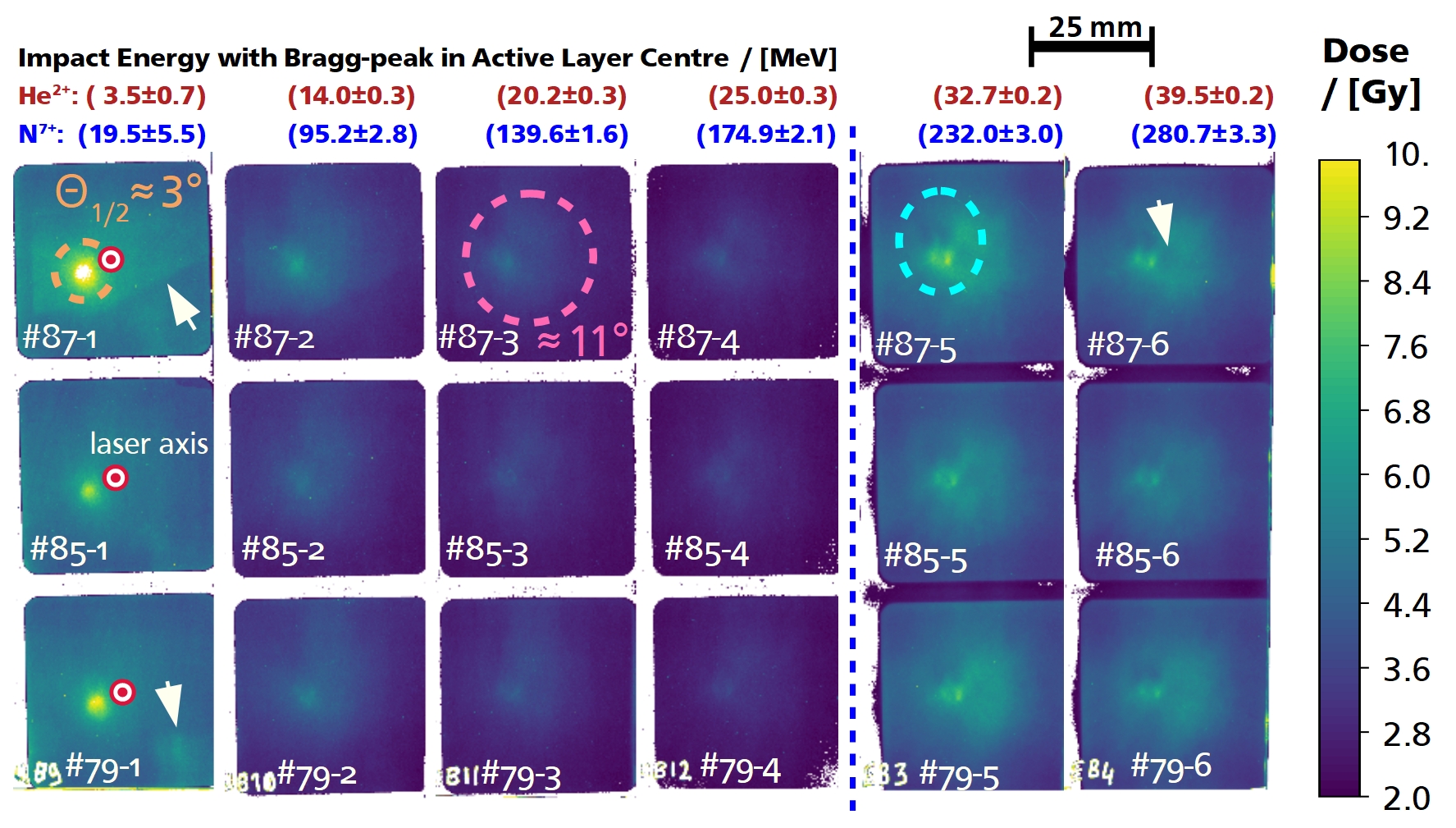}
\caption{Overview on RCF results for shots \#79, \#85 and \#87 converted from RAW to deposited dose maps.}
\label{fig:2018VEGA_RCF_798589_dosemaps}
\end{figure*}

A first qualitative view on the RCF data for shots \#79, \#85 and \#87 \black{with displacement of the nozzle of \SI{0}{\micro\metre}, \SI{-50}{\micro\metre} and \SI{+50}{\micro\metre} respectively, transversely to the main laser axis} is given in figure \ref{fig:2018VEGA_RCF_798589_dosemaps}. For display purpose, data is averaged over the three color channels. The related energies \black{of both possible species} correspond to impact energies of ions with a Bragg-peak in the centre of the active layer.

The laser axis evaluated upon alignment at air pressure is indicated with a white and red crossed marker, it's size represents the alignment uncertainty in air pressure. \black{Photons and particles propagate out of the image plane}. Three features are repetitive and clearly pronounced in all shots. \black{Firstly} a small spot like peak appears aside the aligned laser axis. \black{Its} FWHM half-opening angle is \SI{3}{\degree}, and it is located at approximately \SI{5}{\milli\metre} right of the supposed laser axis and about \SI{1}{\milli\metre} underneath the horizontal plane where is \black{the laser focus}. The sharp peak faints quickly from first to second layer imprint and is only roughly visible in the third layer. The fourth layer does not allow to distinguish the feature. Second, a wide Gaussian peak fills the entire RCF. The FWHM half-opening angle is \SI{11}{\degree}, the maximum position is superposing the laser axis. The Gaussian is visible throughout the RCF stack, from first to last layer. Third, five lobes appear to be clearly visible on the last layers. \black{One appreciates} two maxima and three minima. Both maxima are horizontally aligned next to each other, two minima correspond to these maxima aligned vertically above them. A third minima can be found in proximity of the laser axis, highlighted on the last layer of shot \#87 with a white arrow. Two more features appear on the first layer, also highlighted by white arrows. For shot \#87 there is a shadow visible that demarcates a clear cut from a region of high dose to a region with lower dose. For shot \#79, there is a faint maximum visible at the bottom right corner of the film.

On the first \black{four} layers, the area between dose signals of different shots on the same RCF layer shows no exposure to dose, \black{whereas both last layers \black{of different RCF type} detect dose in \black{these} areas and show overall} higher doses than the \black{first layers}. \black{This indicates that the films used during the experimental campaign where exposed to certain conditions, after their calibration and prior to the experiment, that could have change their response to dose-deposition with respect to the calibrated response.} Other than for the analysis concerning shots with pure Helium jets, we are restraint to the last layer imprints and will regard only the first four layers of the stack. Note, that the wide Gaussian could represent an alpha particle beam\black{, with a similar divergence (11°) as observed in the first study (9°) }.\par

\black{Lobe features} are fainting \black{barely} between third and fourth layer imprint for all color channels, therefore we will assume the fourth layer to be indicative for the background dose delivered by high energy photons and relativistic electrons. \black{For further steps, it} is subtracted as background from all other layers.

\black{The spot like beam feature with low angular divergence shows a steeply decreasing dose throughout the stack. This opens the possibility for either electron or ion projectiles, where possible electron beams could have their maximum dose deposition within the detector filter}.

\begin{figure}
\centering
\includegraphics[width=\columnwidth]{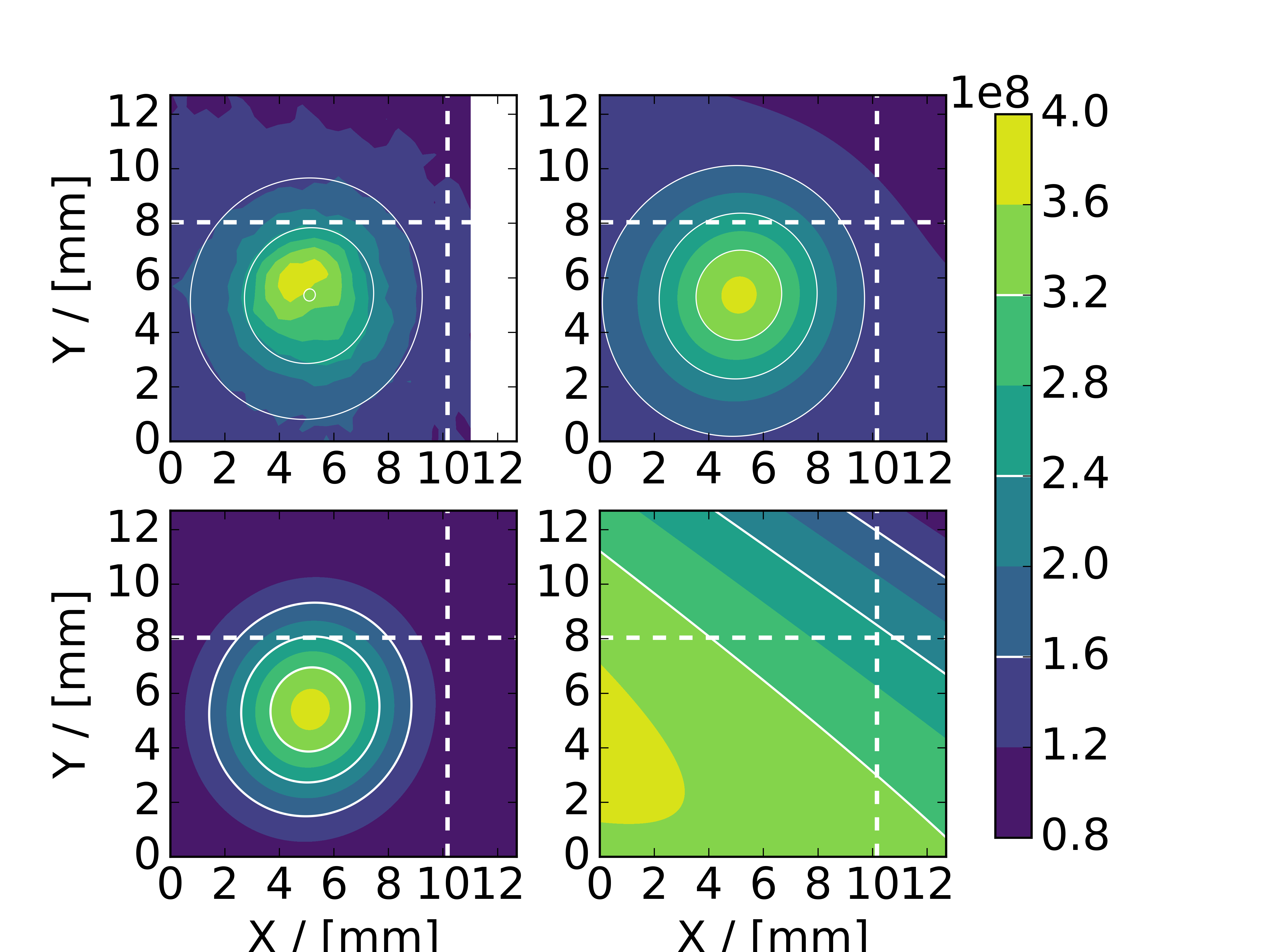}
\includegraphics[width=\columnwidth]{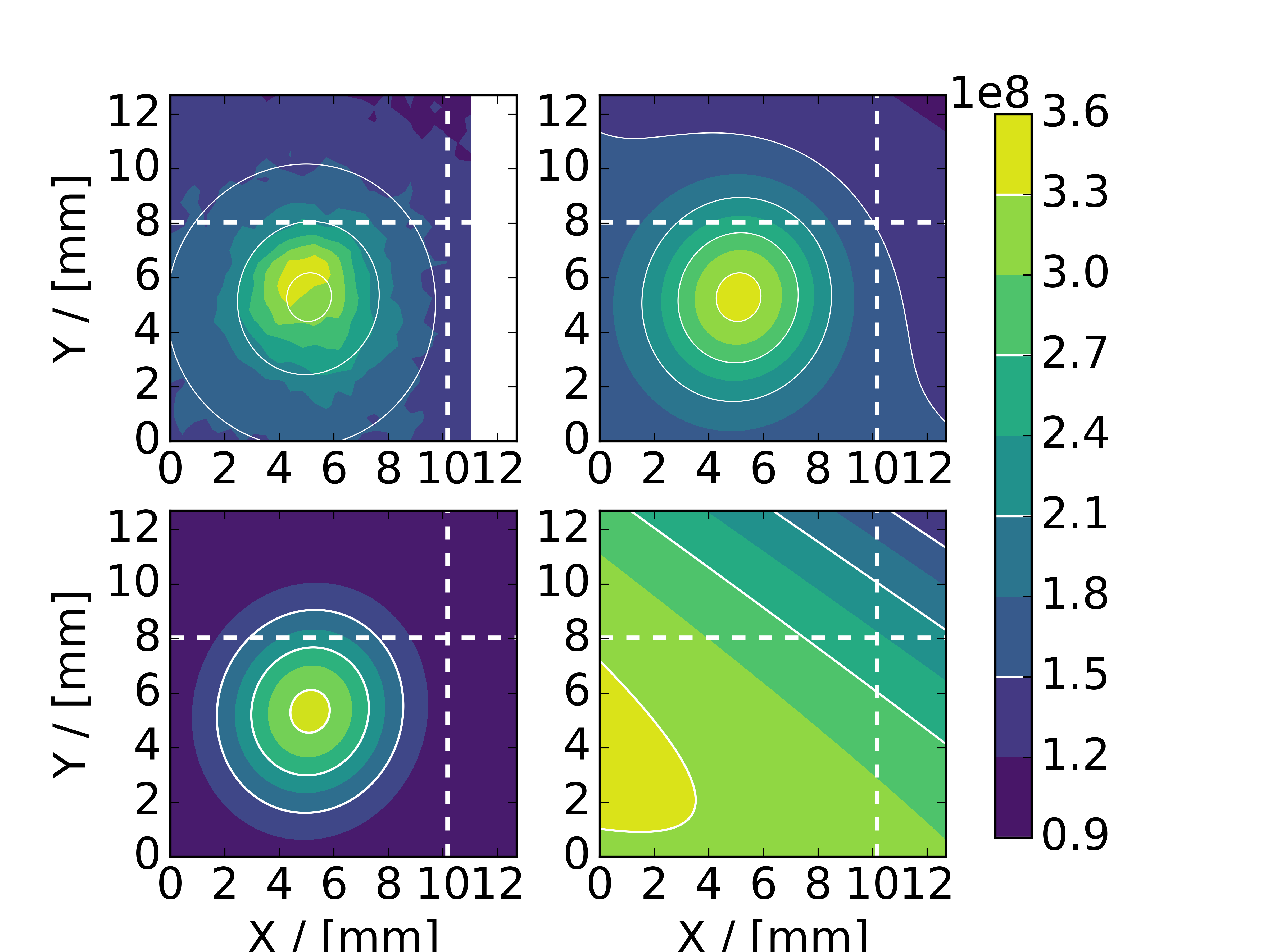}
\includegraphics[width=\columnwidth]{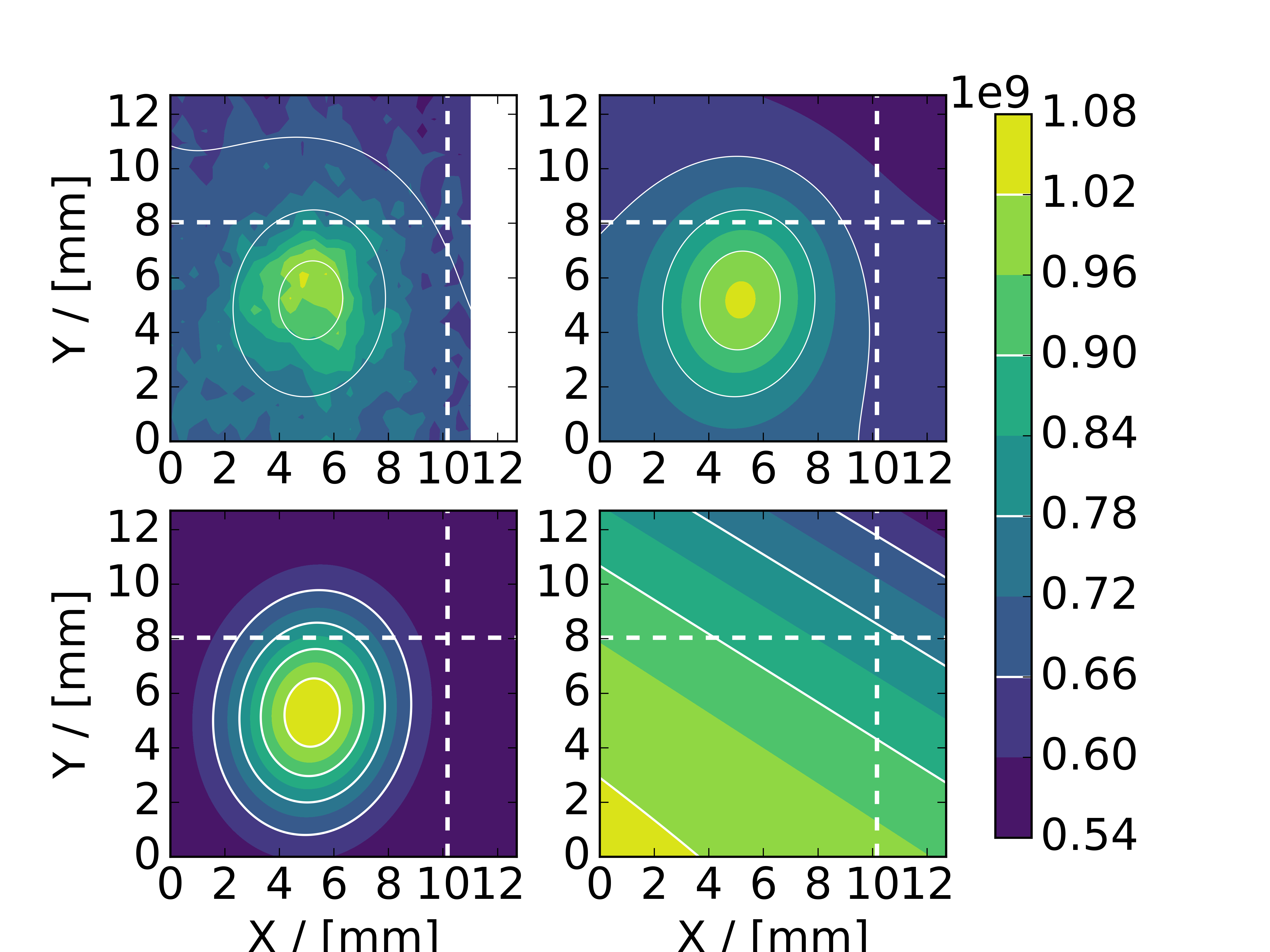}
\caption{Zoom of the peak visible on the first layer U-EBT-3 for shot 79. Each subplot shows one color channel. Therein, clockwise from the top-left corner, (1) the data, (2) a second order polynomial background and a two dimensional gaussian fitted to the data, (4) the gaussian in isolated view and (4) the background isolated. The horizontal plane is parallel to the x-axis and the vertical plane parallel to the y-axis. The color bar is indicative to allow comparison of the four sub-plots with each other. The intersection of white dashed lines indicate the pre-aligned laser axis position.}
\label{fig:2018VEGA_RCF_PLOT_Shot-79_U-EBT-3-1}
\end{figure}

%In a first step...
\black{Presuming alpha particle projectiles as favoured accelerated ion species due to their lower mass, the signal} \black{is fitted} to a 2D-Gaussian superposed with a second order \black{polynomial representing remaining background}, \black{as before.} An example result is plotted in figure \ref{fig:2018VEGA_RCF_PLOT_Shot-79_U-EBT-3-1}.

\begin{figure}[htb]
\centering
\includegraphics[width=\columnwidth]{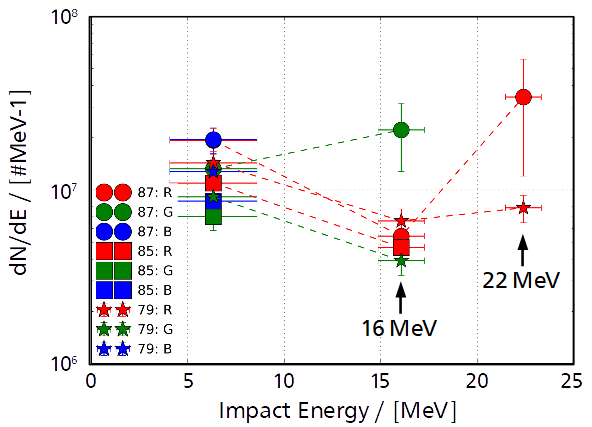}
\caption{\black{Projectile spectrum for three consecutive shots varying the gas jet density profile and transverse position, presuming an alpha particle beam impacting on the stack of RCF.  RCF color channels R, G and B are independently analyzed.}}
\label{RCF_79_85_87_spectrum}
\end{figure}

We see the three color channels of the number density transformed data in the upper left corners of each sub-plot. The upper right corner displays the best fit result, which is then separated into background in the lower right corner and peak in the lower left corner of the plot panel. We appreciate a \black{low level} background signal that is close to a linear evolution, fainting from left to right and bottom to top. The dashed white lines intersect on the laser axis. The background is not symmetric with respect to the laser axis. The \black{Gaussian} fit allows to retrieve the total amount of particles per energy by integration over the full beam. \black{Results} are \black{shown} in figure \ref{RCF_79_85_87_spectrum}, \black{points with} large uncertainties need a critical interpretation, they mostly are of fits on data dominated by the background signal.

The total particle number spectra are flat over the large energy range from \SIrange{6.3}{22.4}{\mega\electronvolt}. With an average of \SI{1E7}{\per\mega\electronvolt} alpha particles, the beams represent bunches of the order of $10^8$ ions. The \black{particle number density} decreases for increasing particle energies throughout the data. The spread of results for different color channels in one shot and the same layer is of the order of the shot-to-shot variation. Nevertheless, the largest amplitudes in this shot series are consistently recorded for shot \#87, the lowest for shot \#85.

%TODO: Inject here analysis aiming at electrons: There are more bulk electrons, so why not having a pronounced electron beam that was not present earlier.

\black{The molecular density of the gas jet in the respective laser shots is \black{shown} in figure \ref{gas_79_85_87_overview}, aside the on-shot interferograms overlaid by line-outs of the streaked self emission.} \black{For small transverse displacements, the first self emission peak corresponds to an un-driven density of \SI{2e19}{\per\cubic\centi\metre}.}

\black{We record lower particle number densities for negative displacement and higher number density for positive displacement (+50 $\mu m$) comparing shots \#85 and \#87. \black{With respect to RCF results, a longer plasma channel and a higher first self-emission peak yield lower particle number densities.} The cut-off energy remains the same, but drops underneath the detection limit allowing a confident fitting.}

\begin{figure}%[ht]
\centering
\includegraphics[width=\columnwidth]{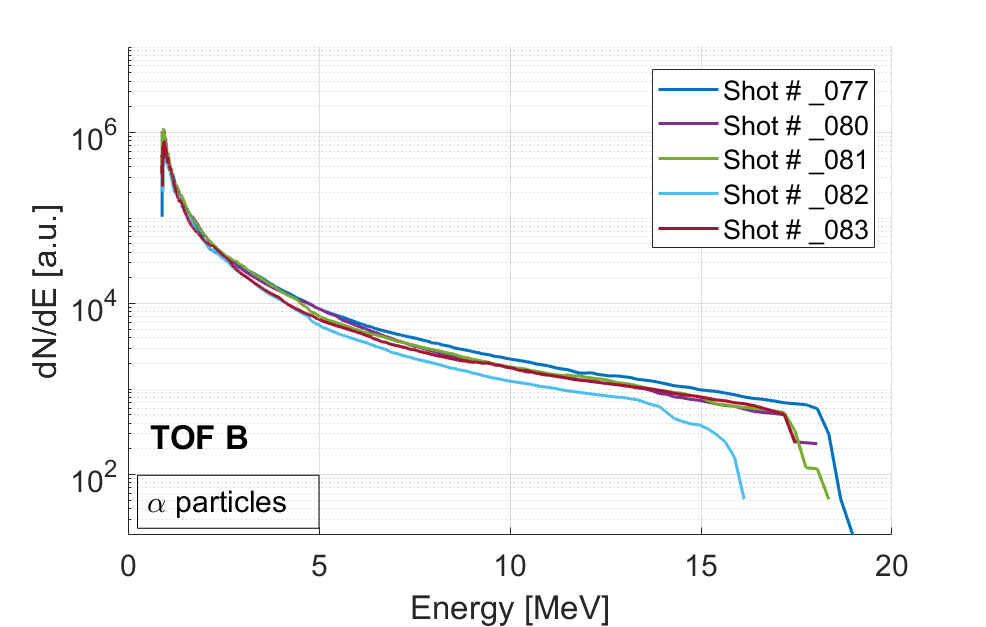}
\includegraphics[width=\columnwidth]{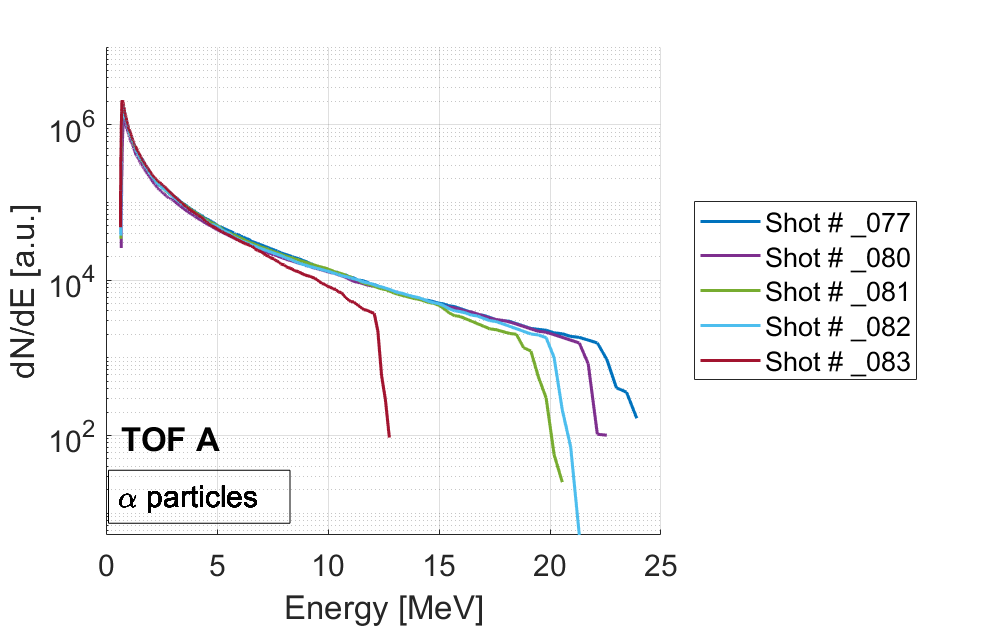}
\caption{Spectra obtained after the analysis of the time-of-flight A (top) and B (bottom) signals assuming alpha particles acceleration for five different shots.}
\label{spectra_alpha}
\end{figure}

\black{Spectra} from the ToF signal in this shot series, can be seen in Fig. \ref{spectra_alpha}. \black{Alpha particle} energies vary from \SIrange{13}{24}{\mega\electronvolt} for the \black{laser forward} emission (16\si{\degree} from the laser axis) and from \SIrange{16}{19}{\mega\electronvolt} for the \black{transverse} emission (106\si{\degree} from the laser axis). Maximum energy cut-offs are less spread in the \black{transverse} emission. The number of particles that arrived to the front detector (ToF A) is slightly lower than the number of particles arriving to the \black{transverse} detector (ToF B).

\black{The corresponding interferometric results and gas density profiles are shown in figure \ref{gas_77_80_81_82_83}. The bulk gas shows only small shot-to-shot differences. The first self emission peak moves towards the focus plane for larger transverse displacements, likely to raise at a fixed density as observed in the first study.} 

\black{The only pair of shots with inter-comparable gas un-driven density profiles \#80 -- \#81 shows similar transverse cut-off energies and number densities. However, we encounter a higher cut-off energy in the laser forward direction in shot \#80. Alike for shots with RCF detector, shot \#80 with shorter plasma channel and more dilute first self-emission peak corresponds to the shot with slightly higher number density in the forward direction.}

\black{The transverse displacement of the gas jet with respect to the laser focus and the shot-to-shot damage to the nozzle do only slightly alter the particle beam spectrum. Accelerated particles may come from the interaction in the bulk gas rather than from the density peak.}

%CR-39 Analysis

\black{In order to determine the ion species and resolve the ambiguity of results obtained with RCF \black{and ToF detectors, we deploy} CR-39.} \black{The chemical etching post-processing unravels clear etch pits on all exposed slaps of CR-39}, an indication for the presence of ion species in laser forward direction. The track density map in the vicinity of the laser axis shows a uniform etch pit distribution over all pit sizes. This may correspond to the large Gaussian distributed \black{dose, visible} with RCF results.

\begin{figure}
\centering
\includegraphics[width=\columnwidth]{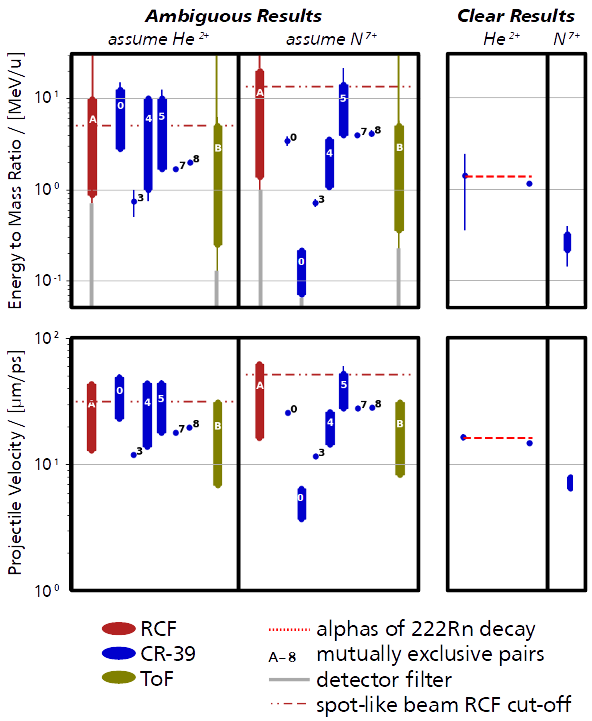}
\caption{Summary of results.}
\label{fig:comp}
\end{figure}

\black{The spectral range of all results is summarized with in Fig. \ref{fig:comp}. Ambiguous are indications for ions of either alpha particles or Nitrogen ions, if both species may yield the same observation. Clear measurements indicate alpha particles of \SI{5}{\mega\electronvolt}, close to the alpha articles issued by the decay cascade of natural Radon gas, as well as low energy Nitrogen projectiles. Some of the ambiguity may be resolved by taking into account the observed range for ions responsible for the wide Gaussian peak on RCF, with FWHM divergence of around \SI{10}{\degree} observed in both series of shots discussed herebefore. None of the etch pits of CR-39 interpreted as Nitrogen stems from energies necessary to penetrate the RCF stack up to the last layers. On the contrary, CR-39 results fully cover the range of alpha energies corresponding to RCF imprints.}

\black{The consistency of spectral ranges for alpha particles found by RCF and CR-39, indicates them to be the ion species responsible for observed large beam features. The ambiguity for small beam features can not be resolved based on the set of diagnostics employed for this study.}

\section{Particle-in-cell (PIC) simulations}
\label{PIC Simulations}
1D particle-in-cell (PIC) simulations have been conducted in order to better understand the experimental results and the complex acceleration mechanisms that could have taken place during the interaction. The parallelized PIC code CALDER developed at CEA was used \cite{Lefebvre2003_NucFus}. We discuss here two of the runs.

The simulated density profile was obtained by fitting a wave-front measurement acquired during the experiment. Since the experimental data presents a lot of fluctuations, likely due to the acquisition diagnostic, a double exponential function was fitted to a gaussian smoothing of the data using a resolution of $0.5\,\mu$m. The exponential fit was extrapolated until the minimum density of $10^{18}\,$cm$^{-3}$. The latter was chosen sufficiently low in order to avoid strong non-physical TNSA that would otherwise appear at the wings of the profile due to a primary interaction of the laser with an initial high target density. Both the density profile and the corresponding exponential fit are plotted in Fig. \ref{fig:setup} (e).

In the following, and unless explicitly marked differently, the density, velocity, time, distance, mass, electron impulsion, ion impulsion, energy and electric field are normalized to the following quantities, respectively:

\begin{equation}
n_c, c, \omega_{0}^{-1}, {c}/{\omega_{0}}, m_e, m_ec, m_ic, m_ec^2, m_e\omega_{0}c/e,   
\end{equation}

\noindent
where $\lambda_{0}$ is taken to be equal to 1 $\mu$m, $n_c$ is the critical density for a 1 $\mu$m laser wavelength equal to $1.11\times 10^{27}\,$m$^{-3}$), $c$ is the velocity of light, $\omega_{0}^{-1}$ is the corresponding inverse laser frequency equal to 5.31$\times 10^{-16}\,$s, and $m_e$ and $m_i$ are the electron and the ion rest masses, respectively.

The main physical inputs of the simulations were a 1D geometry due to the mm spatial scales run over 10 ps, and the impact and field induced ionization of the initially neutral gaseous target, composed by 90$\%$ atomic nitrogen and 10$\%$ helium. Coulomb collisions between all charged particles were considered. Absorbing boundary conditions were used for fields and particles. Each cell of initially neutral Nitrogen and Helium atoms contained 100 particles, while the electrons were created by the ionization processes. The initial ion temperature was $T_i$ = 1 eV. The temporal and spatial resolutions were $0.05\,$fs and $0.016\,\mu$m, respectively. 

The first simulation that was performed shows the interaction of a laser pulse characterized by a normalized laser amplitude $a_0 = 6.8$  (peak intensity $I_L$=$10^{20}$ W/cm$^2$), a wavelength $\lambda_L = 0.8\,\mu$ m and a pulse FWHM duration $\tau_L = 30_,$fs (VEGA II laser pulse characteristics) interacting with the density profile plotted in Fig. \ref{fig:setup} (e), reaching a maximum atomic density $n_{at,max} = 4\times 10^{20}\,$cm$^{-3}$ (0.36 $n_c$). The laser is linearly polarized along the $y-$axis and is injected from the left-side of the simulation box. 

\begin{figure}%[ht]
	\centering
	\includegraphics[width=\linewidth]{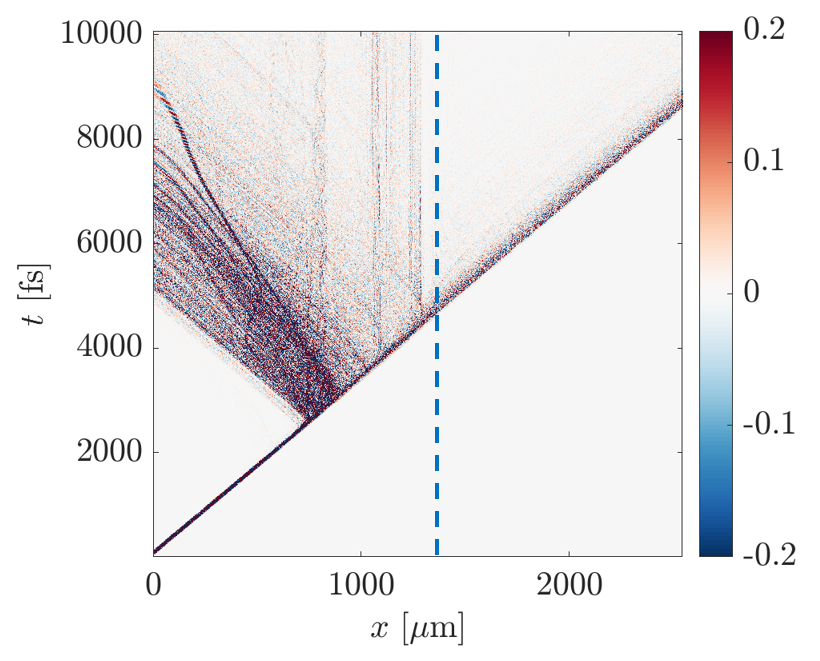}
	\caption{$E_y$ field chart resulting from the simulation with a density profile with $n_{at,max}$ = 4$\times 10^{20}\,$cm$^{-3}$. The blue dashed line marks the density peak.}
	\label{ey_TUCAN}
\end{figure}

Premature laser absorption and reflection in the up-ramp of the gas target (before x = 1000 $\mu\,$m) was observed as seen in the $E_y$ field chart in Fig.\ref{ey_TUCAN}. Density peak at x = 1306 $\mu m$ is indicated by the vertical dashed line.

\begin{figure}%[ht]
	\centering
	\includegraphics[width=\linewidth]{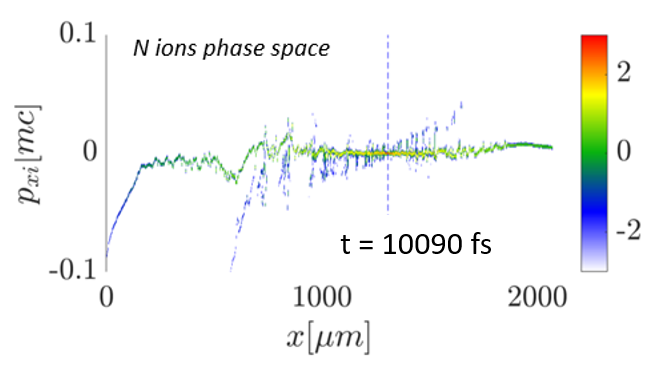}
	\caption{Nitrogen ion phase space in the last time step of the simulation with a density profile characterized by a $n_{at,max}$ = 4$\times 10^{20}\,$cm$^{-3}$.}
	\label{qxpx_sim_1}
\end{figure}

On the final time step of the simulation the ion phase spaces of both Nitrogen ions and alpha particles were very similar: the Nitrogen ions phase space can be seen in Fig. \ref{qxpx_sim_1}. Non-linear ion acoustic waves develop in both up and down density ramps after the laser has crossed. These ion waves depart from a single point into symmetrical, opposite directions, moving as expected at a velocity close to the ion sound speed $c_s$. At some point during their trajectories, reflection of background ions occurs. The reflection in the direction of propagation of the wave is privileged. 

The resultant ion energies in the laser propagation direction are ten times lower than those measured experimentally. One of the possible reasons for this discrepancy is that due to a laser pointing error the laser could have interacted with a less dense target region located on the wings of the profile. The gas target could have also pre-expanded due to the interaction of a $\approx$5 ps-long intensity ramp with a maximum intensity of 10$^{10}$ W/cm$^{2}$, that was systematically seen in the laser pulse contrast measurements performed during the experimental campaign (see inset in Fig. \ref{fig:contrast_with_inset}).

\begin{figure*}[htb]
	\centering
	\includegraphics[width=1.0\linewidth]{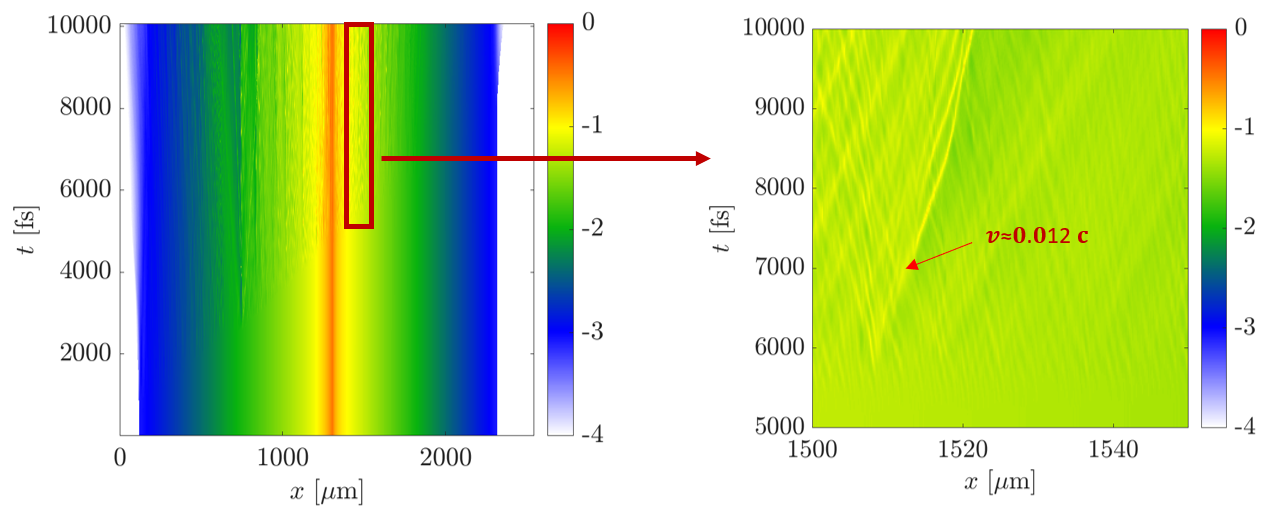}
	\caption{Nitrogen ions density chart (left) and zoom on one of the ion acoustic waves travelling in the down-ramp of the density profile (right) for the first simulation run with $a_0$ = 6.8.}
	\label{qx_TUCAN_and_zoom}
\end{figure*}

The laser is expected to have crossed the profile's density peak as can be assumed by looking at the interferograms of figures \ref{gas_62_63_64_overview},\ref{gas_77_80_81_82_83} and \ref{gas_79_85_87_overview}, where a plasma channel traverses the peak. Therefore, a second PIC simulation with the maximum density reduced by a factor 10 and a slightly increased $a_0$ = 8 was conducted. The maximum experimentally measured $a_0$ was 7.4. The slight increase in the laser intensity could reproduce the effect of laser self-focusing, absent in 1D simulations. 

\begin{figure}
	\centering
	\includegraphics[width=\columnwidth]{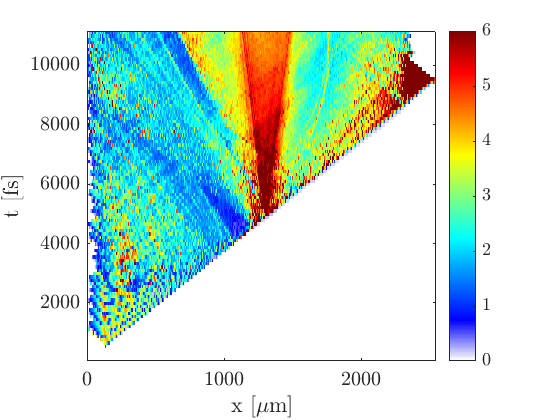}
	\caption{Electronic temperature chart in MeV of the $n_{at,max}$ = 4$\times 10^{19}\,$cm$^{-3}$ (0.036 $n_c$) simulation.}
	\label{Te_TUCAN}
\end{figure}

\begin{figure}
	\centering
	\includegraphics[width=\columnwidth]{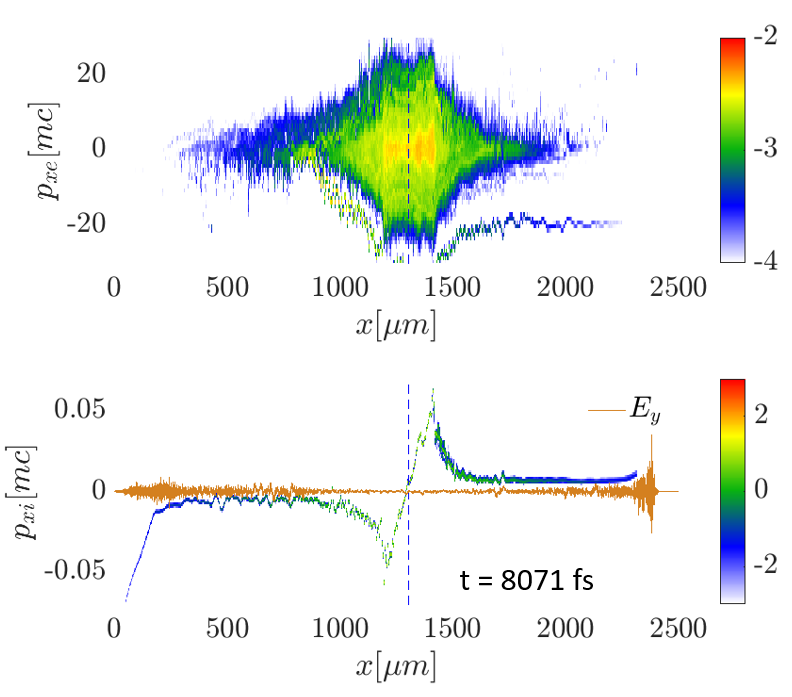}
	\caption{Electrons (top) and nitrogen ions (bottom) phase spaces after the laser has crossed the density peak and traversed the density down-ramp.}
	\label{qxpx_TUCAN_a0_10_t8071fs}
\end{figure}

\begin{figure}
	\centering
	\includegraphics[width=\columnwidth]{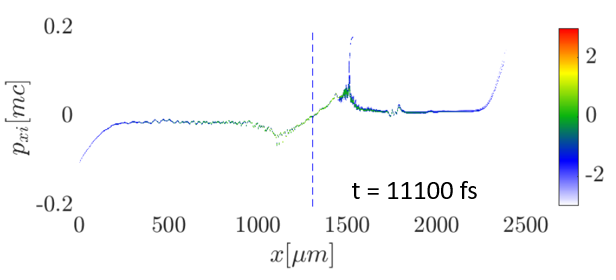}
	\caption{Nitrogen ions phase spaces at the last time step of the simulation with $n_{at,max}$ = 4$\times 10^{19}\,$cm$^{-3}$ (0.036 $n_c$).}
	\label{qxpx_TUCAN_a0_10_t11100fs}
\end{figure}

The ion acceleration processes are different in the conditions of the second simulation. The main mechanism seems to be CSA developed from strong electron pressure gradients in the density peak as illustrated by the electron phase space of Fig. \ref{qxpx_TUCAN_a0_10_t8071fs}, top graph. TNSA (amplified by the profiles abrupt unrealistic density cut) is also seen in the extremes of the density profile in Fig. \ref{qxpx_TUCAN_a0_10_t11100fs} at $x < 200 \ \mu m$ and $x > 2400 \ \mu m$. As has been previously studied \cite{Debayle_2017}, the electronic temperature chart in Fig. \ref{Te_TUCAN} shows that near-critical interaction entails volumetric heating up to relativistic temperatures (normalized $T_e \approx a_0$) through wavebreaking of the electron plasma wakefield and phase mixing between trapped and return-current electrons. Once the laser crosses the density peak the electrons are reflected by the charge separation field created in the down-ramp. An initial single electron vortex located at the density peak and confined by the shock electrostatic barriers evolves into two separate electrostic structures that descend through both the up and down density ramps of the profile, triggering ion acceleration up to $v_i = 0.05 c$, as seen in figure \ref{qxpx_TUCAN_a0_10_t8071fs}. By t = 8070 fs the shock borders are located at $1200\,\mu$m and $1400\,\mu$m. The accelerated ion velocity profile continually steepens while moving down the target ramps and reflects background ions if the correct shock reflection conditions are achieved. Locating ourselves in the frame of the shock, reflection will be produced when the kinetic energy of the incoming particles is higher than the electrostatic potential of the shock barrier. If the ions kinetic energy is lower than the potential they will cross the shock region and suffer deceleration in the process. By the end of the simulation (see ion phase space of figure \ref{qxpx_TUCAN_a0_10_t11100fs}) the reflected ions have been accelerated up to velocities $v_i = 0.15 c$ equivalent to ion energies of $\epsilon_{\alpha}$ = 40 MeV and $\epsilon_{N}$ = 150 MeV. These values are in agreement with the experimentally measured ones. The perturbation in the ion phase space seen at x = 1800 $\,\mu$m is the result the evolution of an initial resonance between the laser frequency and the plasma ($\omega_L = \omega_{pe}$) at x $\approx$ 1320 $\,\mu$m immediately after the laser crosses the density peak (t = 4630 $\,$fs).

% -----------------------------------------------------------------------------------------------------------------------------------------------------------------------------

\section{Conclusion}
\label{Conclusion}

In this experiment we have demonstrated the effective acceleration of broad-spectrum Helium ions beams with cut-off energies above 25 MeV and peaked ion number density $10^8$ MeV$^{-1}$, resulting from the interaction of ultrashort ultraintense laser pulses with near-critical plasmas generated by high pressure gaseous shocked jets. Moreover, this kind of gas jet is capable of delivering the target at HHR, following the available repetition rate of the new generation of high intensity Ti:Sa laser systems (1-10 Hz), yielding therefore an enhanced time-averaged flux of accelerated particles, necessary for most of the medical and industrial applications, such as radiopharmacy. Most of the applications would exploit the unique properties of the laser-driven ion beams (high density, ultra-low emittance, compactness), which may greatly surpass those attained on more costly and larger-scale radiofrequency accelerators.

\black{The acceleration from the shocked} gas jet target shows repetitive results even for small variations in the gas jet profile. \black{The advent of sub-\si{\micro\metre} high precision 3D-printing of mesostructure ceramics \cite{Jo2019} may be beneficial to mass-produce more robust and more precisely tailored shock nozzles. Note that efforts towards automatized nozzle exchange systems and good vacuum systems remain non-trivial technical challenges. Other technical difficulty towards HRR operation of the acceleration platform concerns the vacuum pressuring capability and the strategies to protect the facility compression optics in case of hazardous gas leakages.}

A PIC simulation of the interaction of an $a_0$ = 6.8, $\tau_L = $30 fs, $\lambda_L = 0.8 \mu m$ laser pulse (VEGA II characteristics) with a density profile extracted from experimental measurements (with $n_{at,max} = 4 \times 10^{20} W/cm^2$, 0.36 $n_c$) was performed. Premature laser absorption in the density up-ramp inhibited strong particle acceleration. The accelerated ion energies were 10 times lower than the ones measured experimentally. Ions were accelerated by reflection from IAW developed both in the up and down density ramps of the profile. Since the laser is seen crossing the density peak in the on-shot interferograms, a second simulation with a factor 10 reduction of the density ($n_{at,max}$ = 4$\times 10^{19}\,$cm$^{-3}$, 0.036 $n_c$) was performed, with the intention of simulating conditions that resemble what was experimentally seen. An slight increase in $a_0$ from 6.8 to 8 (the highest experimentally measured $a_0$ was 7.4) was also included in the second run. An increasing $a_0$ is a standard effect of laser self-focusing which is absent in 1D simulations. In this case, ion acceleration through collisionless shock formation having its origin in strong electron pressure gradients located in the density peak was observed. TNSA acceleration was also seen in the target extremes. Strong electron heating up to $T_e \approx a_0$ was obtained. The final reflected ion velocities are in agreement with the experimental measurements.

\section{Foreseen applications}
\label{Applications}

\black{Generalizing to the laser repetition rate, the alpha particle beam demonstrated in this study has a beam-current of \SI{64}{\pico\ampere\per\hertz\per\mega\electronvolt}. Such beam currents are of interest in radio-chemistry. Radiolysis, the breaking of chemical bonds by radiation, is conducted with alpha projectiles with tens of \si{\mega\electronvolt} \cite{Saini1987} and beam currents of \SIrange{10}{100}{\nano\ampere} -- parameters that can be achieved even with low repetition rate. Further optimization of the platform is of interest regarding short ion bunches for picosecond pulse
radiolysis research\cite{Ba2008} and may yield to high brilliance sources for not time averaged observations conducted with single shots \cite{Bo2020}.}

\black{Alpha therapy is a promising approach for clinical oncology, promoting the targeted destruction of a cancerous tumor and metastases in vivo \cite{Ch2018}. This form of radiotherapy is not carried out by external radiation, but by administering radioactive substances to the patient. The cells are reached through the bloodstream. Unlike chemotherapy, the treatment is aimed at diseased cells due to the targeted binding properties of compound molecules and thus reduces possible side effects \cite{IAEA_HHR_15}. Alpha-emitting radionuclides can be produced by interaction of ionizing radiation with a suited target material, which is a major challenge for reasons of the needed quantities \cite{Ch2018}. Astatine-211 is a suited candidate for its intermediate half live time of \SI{7.2}{\hour} and low risk for side effects \cite{Ma2018}, which can be produced in a $(\alpha,2n)$ reaction by irradiating bismuth-209 with alpha particles of \SI{28}{\mega\electronvolt}. Beam currents of several hundred \si{\micro\ampere} are necessary for efficient isotope production and therefore repetition rates of the \si{\kilo\hertz} and alpha particle number densities by three orders of magnitude higher than observed -- a clear motivation for further studies, where more energetic lasres can play an important role.}

% -----------------------------------------------------------------------------------------------------------------------------------------------------------------------------

\section{Acknowledgments}

\black{We thank C.~Sergeant and Ph.~Barberet (CNBG, Univ. Bordeaux, Gradignan) as well as B.~Bondon and all the technical staff (UF Chimie, Univ. Bordeaux, Campus Talence), for their unconditional help and the provision of the necessary analysis equipment and expertise for the etching of CR-39.}

We would like to thank Ashland Specialty Ingredients G.P., that promptly agreed to help us detecting low doses of narrow-band spectra in the low \si{MeV} range by producing a then not available special type of RCF. The unlaminated EBT-3 (U-EBT-3) masters this task with its sensitivity equal to EBT-3 in a thin layout. Many thanks to J\'{e}rome Caron that allowed us to take calibration data for RCF with his support at Institut Bergoni\'{e} in Bordeaux.

This work has received funding from the European Union's Horizon 2020 research and innovation program under grant agreement no 871124 Laserlab-Europe.

We received financial support from the French State, managed by the French National Research Agency (ANR) in the frame of “the Investments for the future” Programme IdEx Bordeaux – LAPHIA (ANR-10-IDEX-03-02).

We acknowledge the support from the LIGHT S\&T Graduate Program (PIA3 Investment fot the Future Program, ANR-17-EURE-0027).

We acknowledge GENCI for providing us access to the supercomputer Irene under the grants no. A0070506129 and no. A0080507594.

This project has received funding from: i) the Spanish Ministerio de Econom\'{i}a y Competitividad through the PALMA Grant No. FIS2016-81056-R, ii) the Spanish Ministerio de Ciencia, Innovaci\'{o}n y Universidades ICTS Equipment grant No. EQC2018-005230-P, iii) the LaserLab Europe V Grant No. 871124 and iv) the Junta de Castilla y Le\'{o}n grants No. 659 CLP087U16 and No. CLP263P20.  

% -----------------------------------------------------------------------------------------------------------------------------------------------------------------------------

\section{Additional Material: methods}\label{sec:methods}

\subsection{Laser Diagnostics}
\label{Additional Material: Laser Diagnostics}
Key to a good focal spot quality was a two translations plus two tilt axis motorization system of the OAP mount allowing for flexible alignment under vacuum. The focal spot was optimized every day at air pressure, yielding a FWHM of 7.5 $\pm$ 1.4 $\mu m$ with a varying percentage between 20\% and 40\% in the FWHM.

The pulse duration was verified employing an autocorrelator, yielding 37.06 $\pm$ 0.81 fs. The temporal contrast was measured with the commercially available spectral filtering third-order cross-correlator system Sequoia (Amplitude Technologies).

\begin{figure}[htb]
	\centering
	\includegraphics[width=\columnwidth]{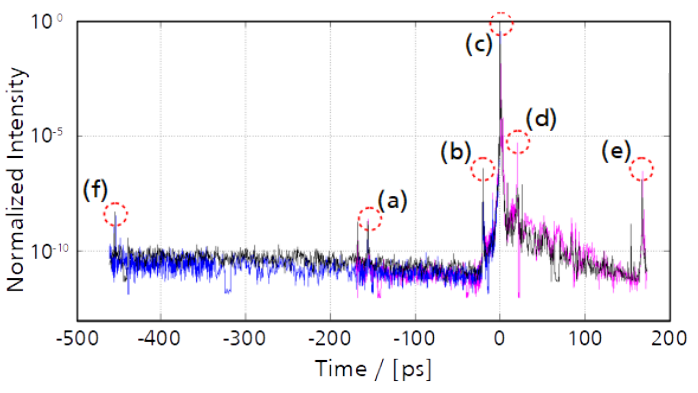}
	\caption{\black{Three independent Sequoia measurements. Identified pre and post-pulses have been marked with a dashed circle.}}
	\label{fig:contrast_appendix}
\end{figure}

Three independent Sequoia measurements of the temporal contrast at best compression with a \numrange{e-10}{e-11} ASE level can be seen in fig. \ref{fig:contrast_appendix}. Identified pre- and post-pulses are located at (a) \SI{-156}{\pico\second} due to the pockels cell at the regenerative amplifier, (b,d) \SI{\pm 20}{\pico\second} as likely artifacts of the measurement itself and (e) \SI{167}{\pico\second} due to the pockels cell at the regenerative amplifier. (f) is unidentified, (c) indicates the main pulse.

\subsection{ToF Analysis from PIN diodes data}\label{sec:material_ToF}

\black{PiN diodes and many other Time-of-Flight (ToF) detectors rely on time resolved voltage signal read-out, e.g. by fast oscilloscopes. Projectiles impacting on ToF detectors alter proportionally the detector voltage.}

\black{The} \black{PiN configuration of the diodes} \black{includes} an extra neutral \black{layer "I"} in the middle of the doped \black{layers "P" and "N", which} allows for a quantum efficiency enhancement \black{by} increasing the \black{cross-section of ionizing radiation} with the diode substrate, \black{and a faster} response \black{in absence of a} minority carriers current \black{that delays} the signal current formation. The \black{temporal resolution is \SI{4.5}{\nano\second}} FWHM. \black{With} pressurized gas entering the vacuum chamber, the photodiode exhibits another advantage \black{with its} bias voltage \black{of} \SI{-60}{\volt}. \black{This is comparatively low to other ToF} detectors \black{such as} microchannel plates, which \black{are} polarized with bias voltages of \black{the order of \si{\kilo\volt} which represent risk of short-circuit currents through the momentaneously imperfect vacuum}.

\begin{figure}%[ht]
\centering
\includegraphics[width=\linewidth]{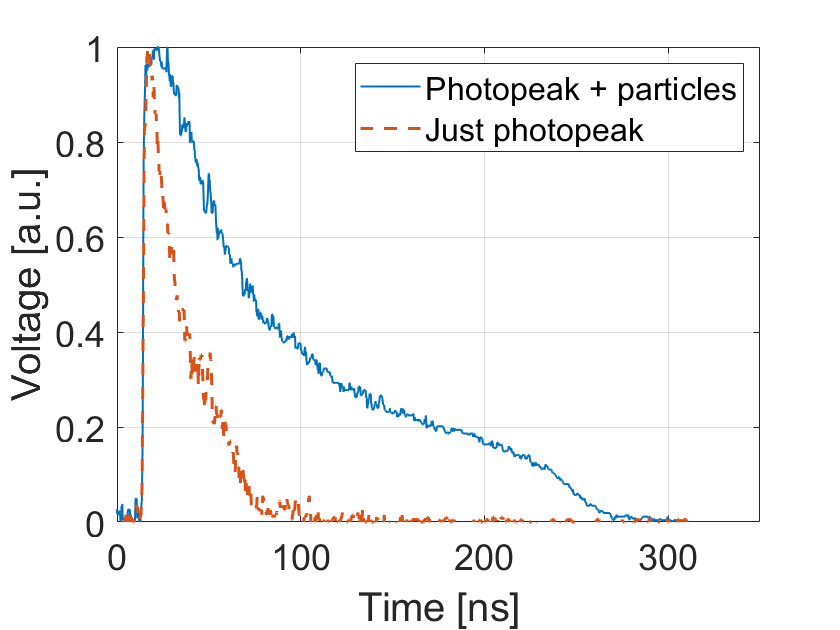}
\caption{Comparison between \black{the detector RAW signal comprising photopeak and all accelerated particle species} (blue full line) and a signal \black{with photopeak only} (orange dashed line).}
\label{comp_photopeak_norm}
\end{figure}

The obtained signal includes the photopeak and the accelerated particles contribution. An example of the obtained signal is plotted in figure \ref{comp_photopeak_norm}. The photopeak refers to the \black{peaked detector response to photons emitted from the laser-plasma interaction and successive cooling. It is pronounced with a} first \black{sharp rising edge} of the signal \black{allowing one} to retrieve \black{a reference detection time $t_0$ for photons travelling at the speed of light $c$. For the charged particle projectiles, we obtain velocities $v_\mathrm{p} = s/t_\mathrm{p}$ corresponding to detection times $t = t_0 + t_\mathrm{p} - s/c$}. Knowing the drifting length $s$ and assuming a paricle mass, one can retrieve an energy spectrum of the accelerated particle species.

\black{In order to distinguish between several possible species, the calculated low energy cut-off can be compared to the cut-off which is defined by the thickness of a filter in front of the detector.} Electrons are \black{generally} present in \black{laser} accelerated species. However, by knowing the filter thickness in front of each photodiode \black{the signals acquired during this campaign can not correspond to electron projectiles}. Therefore, the \black{signal outside the photopeak} must be attributed to ions being accelerated. In our experiments, these could be Helium and/or Nitrogen ions.

During the campaign there were four shots where no particle acceleration was seen in the detector (due to the nature of that specific shot conditions) and only the x-ray photopeak was acquired. The photopeak signal was found to be almost constant in intensity and decay time from shot-to-shot and to follow an exponential decay with a mean $\tau$ constant equal to 0.2 ns. Therefore and in order to obtain the ion part of the signal an exponential decay with the mentioned $\tau$ constant is subtracted from the original signal.

\begin{figure}%[ht]
\centering
\includegraphics[width=\linewidth]{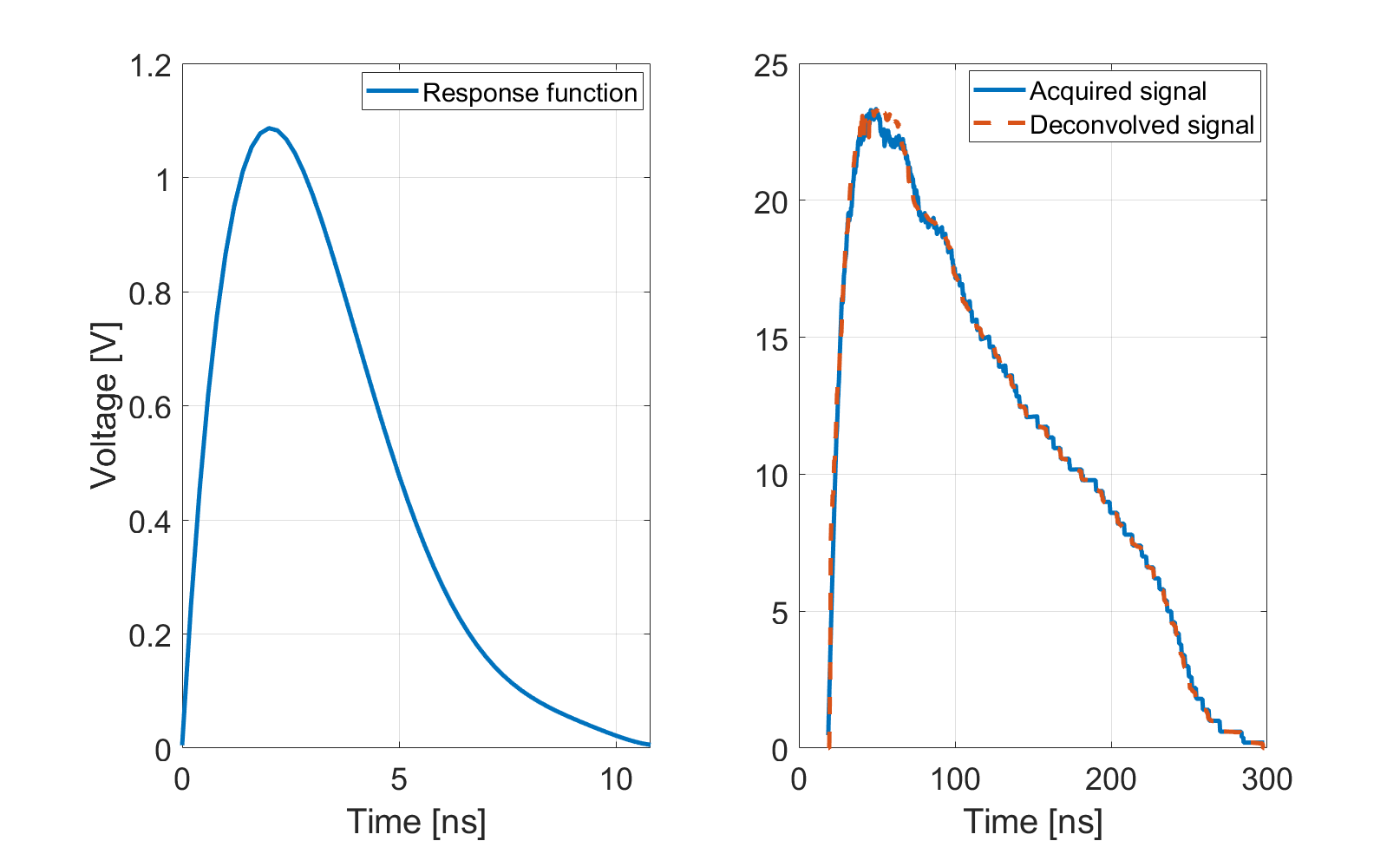}
\caption{Photodiode detector impulse response (left) and signal after deconvolution (right).}
\label{deconv_subplot}
\end{figure}

Once the photopeak contribution has been subtracted the original signal must be obtained by performing a deconvolution with the photodiode's impulse response, which is an exponentially modified gaussian. This was characterized in a previous campaign at CLPU by shooting low energy laser pulses (considered as an impulse input signal) directly at the detector. It was seen to be independent from the applied bias voltage. The detectors response function and the deconvolution result can be seen in fig. \ref{deconv_subplot}.

\black{To our best knowledge} available calibration data \cite{corallo1980x} is related to X-ray energy deposition \black{with a} linear calibration coefficient of \SI{0.282}{\joule\per\coulomb}.  However, since the photodiode response can be assumed to be linear if no saturation is produced, the \black{ordinate} can be used in a relative manner \black{representing relative number density counts: see Fig. \ref{spectra_alpha_61_62_64_66} and Fig. \ref{spectra_alpha}}.

\subsection{Passive particledDetectors}

Both passive detector types, RCF and CR39, are \black{used in alternation and therefore} clamped \black{in a fully motorized holder. The casing comprises} two plates with a multitude of squared frames of \SI{40}{\milli\metre} sides. The recesses are cut out of \SI{3}{\milli\metre} thick Aluminium plates. The detector array \black{was displaced} behind a \SI{5}{\milli\metre} thick aluminum plate acting as \black{radiation-}shielding\black{, only with} a squared opening of \SI{40}{\milli\metre} times \SI{40}{\milli\metre} centred around the aligned laser axis.

\subsubsection{RCF data Analysis}\label{sec:material_RCF}

RCF self-develop under the influence of ionizing radiation and show a change of color (darkening). A key feature is that the changes in the absorption spectrum of a film are bijective with respect to the received dose $D$. The films contain sub-\si{\micro\metre} sized chromophore components that allow a very good spatial localization of the impact of projectiles after radio-synthesis. This means that a space-resolved measurement of the scanning wavelength-dependent optical density $\iota (\lambda , \vec{x} )$ can be transformed into a localized dose measurement $D(\vec{x})$. The received total dose is proportional to the so called linear energy transfer (LET) from single projectile--stack interactions. For ion beams, it is possible to retrieve both spectrum and phase space of an ion beam of a known species as the LET in solid state matter is well quantified \cite{Zi1999}.

The particularity of ion stopping with respect to electron- and photon-stopping is a maximum of the stopping power for low ion energies with small scattering cross-sections: the stopping power maximum is always located at the end of the penetration range. It is called Bragg-peak \cite{Si2006} and accurately describes the LET of ions in a stack. The film response to ions can be easily distinguished from high-energy background electrons, that may even traverse the stack, as their imprints faint only slowly with increasing depth. Thus the cut-off energy of the ion spectrum $E^\text{\tiny max}$ can be estimated from the energy $E^\text{\tiny out}$ for which the stopping range corresponds to the depth of the last colored film in the stack. From this energy on, it is possible to de-convolute the particle number density spectrum with respect to projectiles of different energies going from the last imprinted layer to the first imprinted layer in the stack, from high to low energies in the spectrum \cite{Nue2009}.

\begin{figure}
\centering
\includegraphics[width=\columnwidth]{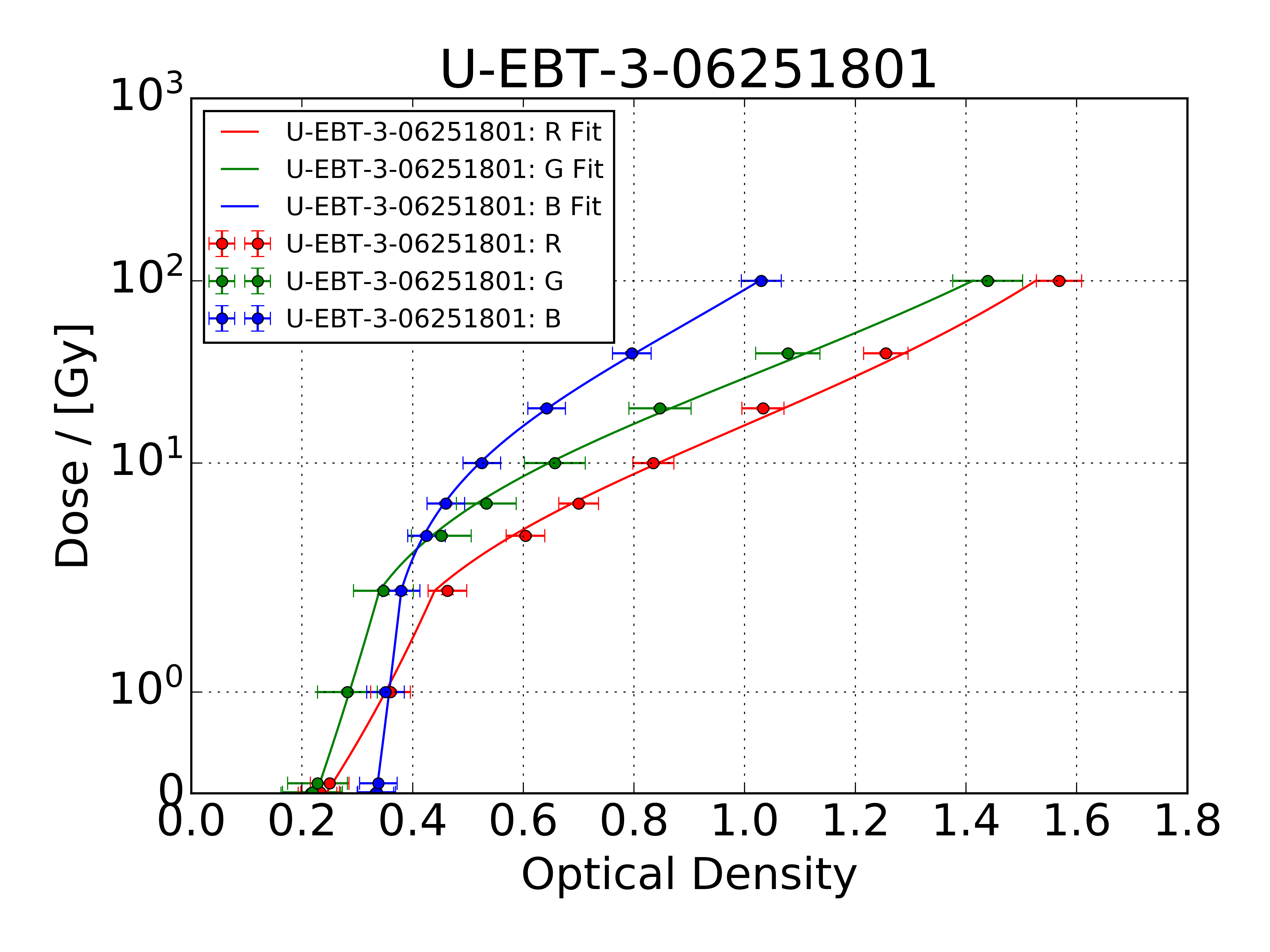}
%\verbatiminput{images/plotter_U-EBT-3-06251801_R.dat}
%\verbatiminput{images/plotter_U-EBT-3-06251801_G.dat}
%\verbatiminput{images/plotter_U-EBT-3-06251801_B.dat}
\includegraphics[width=\columnwidth]{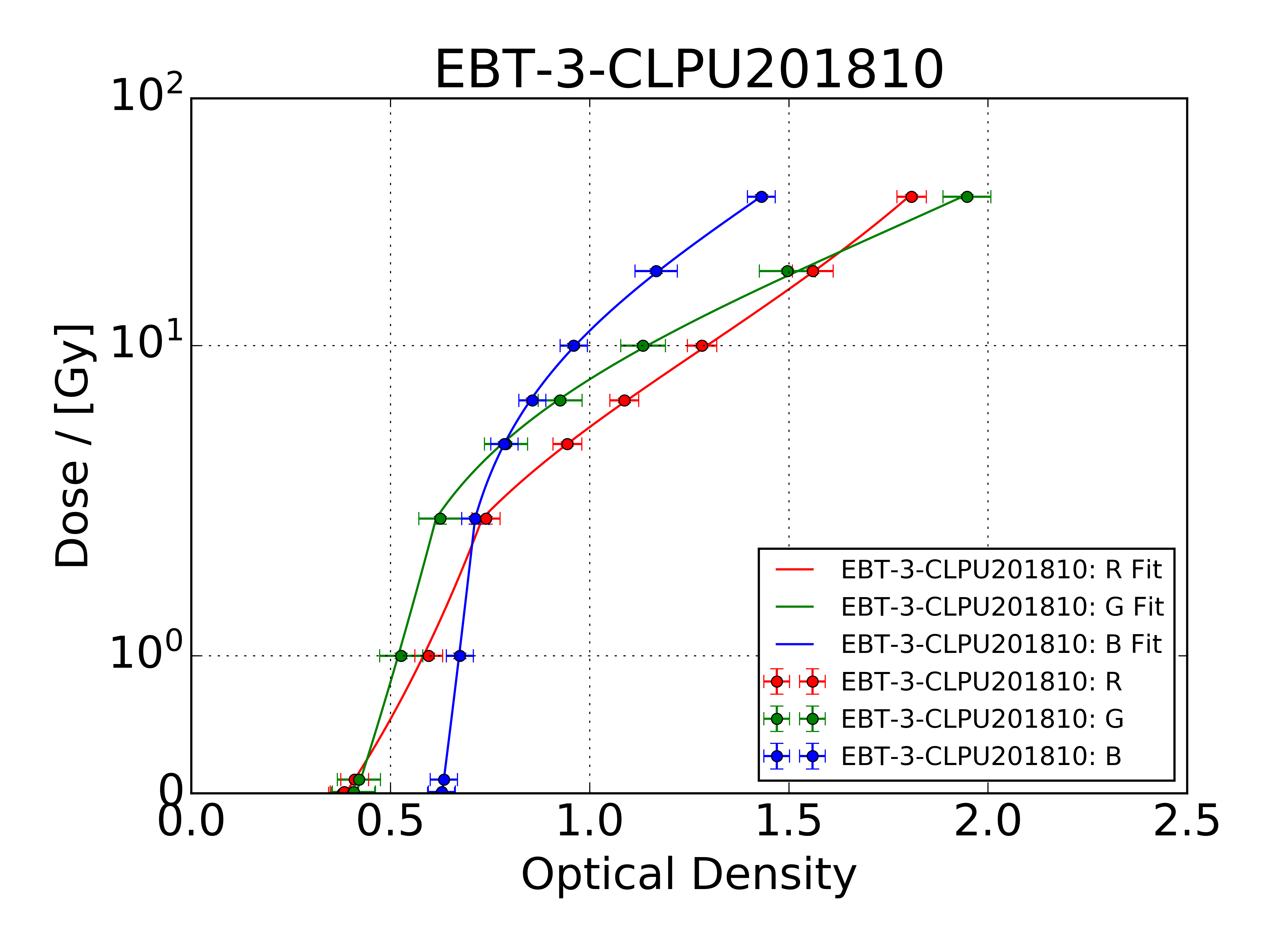}
%\verbatiminput{images/plotter_EBT-3-CLPU201810_R.dat}
%\verbatiminput{images/plotter_EBT-3-CLPU201810_G.dat}
%\verbatiminput{images/plotter_EBT-3-CLPU201810_B.dat}
\caption{Channel-wise fit of the universal dose response model to calibration data for U-EBT-3 batch number 06251801 (top) and EBT-3 batch number 10251701 (bottom). The dose deposition for calibration data is performed at a medical accelerator unit for clinical oncology based on electron- and photon irradiation at Institut Bergoni\'{e}, Bordeaux. The dose precision better than \SI{1}{\percent}. The dose is delivered in bunches of \SI{1}{\centi\gray} with a dose rate of \SI{10}{\gray\per\minute} with a macro-bunch length maximum of \SI{50}{\gray}. Macro-bunches are intersected by a mandatory operator request that takes approximately \SI{10}{\second} for performance in average. Scans are acquired with the EPSON EXPRESSION11000XL flat-bed scanner that comprises a cold fluorescent Xenon light source for illumination.}
\label{fig:dose-OD-U-EBT-3-06251801-EBT-3-10251701}
\end{figure}

Our data is converted to dose maps with calibration shown in Fig. \ref{fig:dose-OD-U-EBT-3-06251801-EBT-3-10251701} obtained at a calibrated medical accelerator unit. Fits employ the universal dose response model \cite{Ma2015}
\begin{align}
\iota_\text{\tiny ch} &= A_\text{\tiny ch}^\text{\tiny fit} \cdot \left( 1 - (1+\frac{D}{k_\text{\tiny ch}^\text{\tiny fit}})^{\theta} \right) + \iota_{0,\text{\tiny ch}}^\text{\tiny fit} \quad .
\end{align}

The sensitometric curve for the optical density of each color channel $\iota_\text{\tiny ch}$ is a function of dose $D$. The fit parameters $A_\text{\tiny ch}^\text{\tiny fit}$ and $k_\text{\tiny ch}^\text{\tiny fit}$ depend on channel and scanner system, the off-set $\iota_{0,\text{\tiny ch}}^\text{\tiny fit}$ depends on the life-cycle of the non-irradiated film and $\theta = 0.436 \pm 0.005$ \cite{Ma2015} represents a universal constant that depends on the sensitive material.

\subsubsection{CR-39 data analysis}\label{sec:material_CR39}

CR-39 are made of polyallyl diglycol carbonate polymers with diethylene glycol bis(allyl carbonate) monomer $\text{C}_12\text{H}_{18}\text{O}_7$. During exposure to ionizing radiation, projectiles collide with the polymer chains and some of them break \cite{SP1986,Ma1987,Th1995,Ch1997}. As a result, the projectile path is permanently imprinted in the plastic with disrupted chains, forming a latent track, and partially damaged chains, forming radicals. Projectiles with a low LET from \SIrange{10}{15}{\kilo\electronvolt\per\micro\metre} do not lead to the formation of tracks \cite{Ha2008}. Thus, single electrons do not produce CR-39 track formation, only electrons in high doses of several \si{\kilo\gray} trigger radical formation \cite{Ch1988}.

Etching the CR-39 with an alkali solution yields ablation of surface material due to reactions of the polymer with the hydroxide ions\cite{SP1986}. The material ablation rate is constant for a pure homogeneous plastic. For exposed regions along tracks with already broken links, the ablation rate is higher and a crater develops, often referred to as etch pit.

Main observables of etch pits at the detector surface are depth, diameter and their evolution with etching time. Observables are bijective with a particular pair of ablation rates and allow to retrieve them along the full projectile track. Such mapping can be compared to calibration data in order to find likely ion species and impact energy ranges. Additionally, the direction of projectiles can be retrieved from the inclination of the crater axis to the detector surface and the ellipticity of the crater.

TasTrak CR-39 Plastic Sheets \cite{TASTRAKMSDS} \black{used for this work are reported to show} a good inter-comparability of calibrations \cite{Zh2019} \black{to} CR-39 of Page Mouldings (Pershore) Limited. This is of further importance as we rely on third-party calibration data \cite{Az2012,Az2013} acquired with CR-39 of Page Mouldings (Pershore) Limited. 

We employ $\text{Na}\text{O}\text{H}$ solution of $6.25~\text{N}$ in distilled water for etching the CR-39 from all sides in a bath at constant temperature of \SI{70}{\celsius}. For specific etching times in a range from \SI{2}{\minute} to \SI{15}{\hour}, we take a series of \SI{447.63}{\micro\metre} $\times$ \SI{335.4}{\micro\metre} images acquired in $12\text{~bit}$ with a fully motorized CarlZeiss microscope with approximately \SI{2}{\micro\metre} resolution.

\begin{figure}
\centering
\includegraphics[width=\columnwidth]{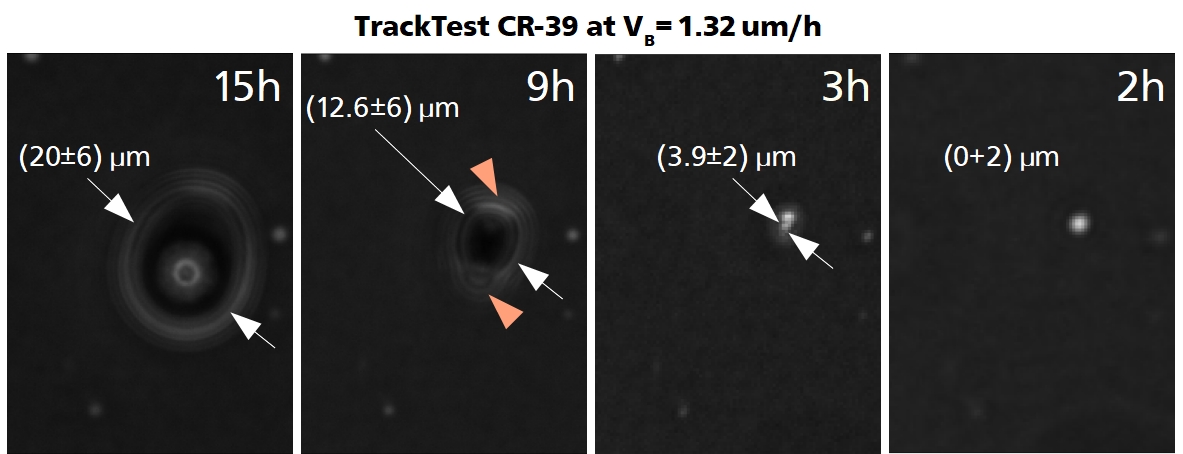}
\caption{Diameter evolution of an individual etch pit on a TasTrak CR-39 with strioscopic snapshots at \SI{2}{\hour}, \SI{3}{\hour}, \SI{9}{\hour} and \SI{15}{\hour} of etching with $6.25\text{~N}$ etchant at \SI{70}{\celsius}. The diameter evolves considerably and is not detected for \SI{2}{\hour} of etching as it may be underneath the resolution of the microscope system.}
\label{fig:CR39_diameterresults_centreionrcf_individual}
\end{figure}

\black{We base a first analysis} on the pit diameter evolution with etching time. We select one particular approximately \SI{20.}{\micro\metre} large etch pit in vicinity of the presumed laser axis at \SI{15}{\hour} and follow it back in time. Figure \ref{fig:CR39_diameterresults_centreionrcf_individual} shows strioscopic microscope images with the growing pit. Also imaged are several small craters that do not show large growth rates.

The diameter evolution of the large pit corresponds well to the expected evolution for alpha particles of \SI{5.4+-4.2}{\mega\electronvolt} within the margin of the measurement uncertainty, especially due to a diameter underneath the detection limit for \SI{2}{\hour} of etching. In the same figure, small etch pits of \SI{10+-5}{\micro\metre} are visible that are present even for the earliest time, which can not be related to alpha particles. If they belonged to low energy alpha particles, they were expected to grow with about the same growth rate as the finally large etch pit. If they belonged to high energy alpha particles, they should not be visible. Their size does not match any alpha energy nor any energy of Nitrogen projectiles. \black{They could be related to detector material impurities.}

\begin{figure*}
\centering
\includegraphics[width=\textwidth]{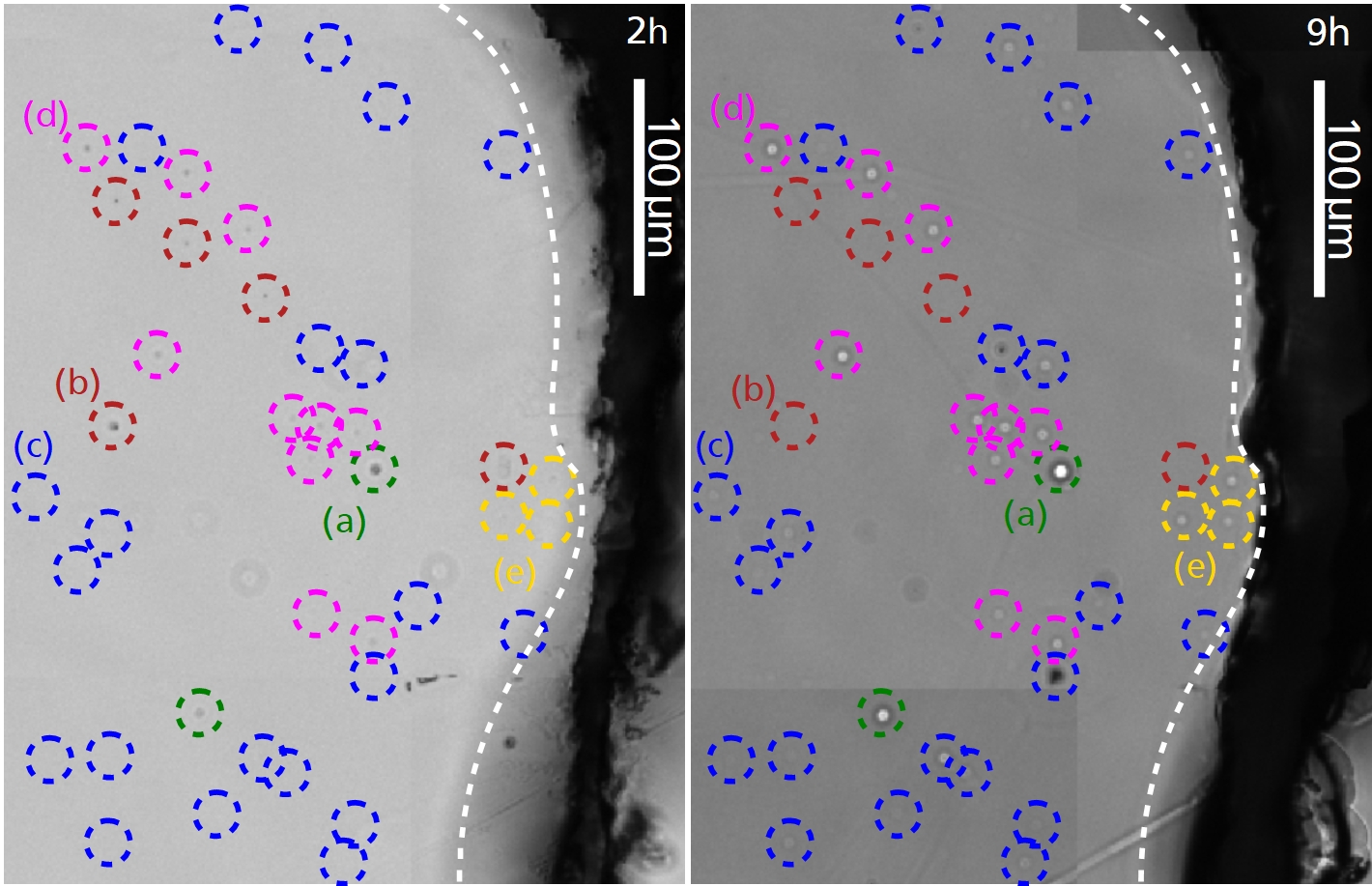}
\caption{Diameter evolution for similar groups of etch pits on a TasTrak CR-39 with snapshots at \SI{2}{\hour}, and \SI{9}{\hour} of etching with $6.25\text{~N}$ etchant at \SI{70}{\celsius}. Highlighted are groups (a,b,d,e) with circles concentric to the etch pit position for the short etching time and (c) with circles concentric to the pit position for long etching time. The dashed line delimits a scratched area, we do not see the edge of the plastic sheet.}
\label{fig:CR39_diameterresults_monster_individual}
\end{figure*}

\black{In order to confirm the appearance of alpha particles,} we choose another typical area far from the laser axis to investigate etch pits with diameters at the detection threshold, see figure \ref{fig:CR39_diameterresults_monster_individual}. The respective dashed circles indicate all etch pits in the field of view, other circular features are microscope artifacts. The circles are located at the same position for both images taken after \SI{2}{\hour} and \SI{9}{\hour} of etching.

We see that etch pits grow inclined with the surface, their centre shifts with an inclination that is pointing to the laser axis for most pits. Group (a) with already large pits with \SI{5.8+-0.9}{\micro\metre} diameters for \SI{2}{\hour} and slow growth to \SI{6.8+-1.9}{\micro\metre} for \SI{9}{\hour} follows an evolution that suits neither nitrogen nor alpha projectiles. We see craters (b) of \SI{9+-6}{\micro\metre} that are visible for short but not for long etching, which are most likely shallow surface artefacts. Inversely, the most populated group of pits (c) is underneath the detection limit for short etching and reveals pits with diameters from \SIrange{6+-2}{13+-3}{\micro\metre} for \SI{9}{\hour} of etching. Such pits suit either alpha particles with energies from \SIrange{4}{40}{\mega\electronvolt} or nitrogen within an energy range of \SIrange{15}{50}{\mega\electronvolt}. Grouped with (d) are pits that are at the detection limit for \SI{2}{\hour} with diameters of \SIrange{2+-2}{5+-2}{\micro\metre}. The pits grow to \SIrange{5.8+-1}{7.4+-2}{\micro\metre} after \SI{9}{\hour}. Where diameters are well above the microscope resolution, the craters must be very shallow to show a fainting image. Such pits suit nitrogen within an energy range of \SIrange{1.2}{2}{\mega\electronvolt}. Furthermore we see tracks with coherent but different inclination (e), not coming from the presumed laser axis.

It is noteworthy that alpha energies in the \si{\mega\electronvolt} range are typical for alpha particles from the decay of natural Radon gas and its chain of decay. $^{222}\text{Rn}$ has a half life time of \SI{3.824}{\day} and splits into alpha particles of \SI{5.49}{\mega\electronvolt} and $^{218}\text{Po}$. The pits grouped in (e) may be indeed issued by a chain of radon decay within \si{\centi\metre} in the vicinity of the CR-39.

\begin{figure*}
\centering
\includegraphics[trim={0 0 0 5mm},clip,width=\textwidth]{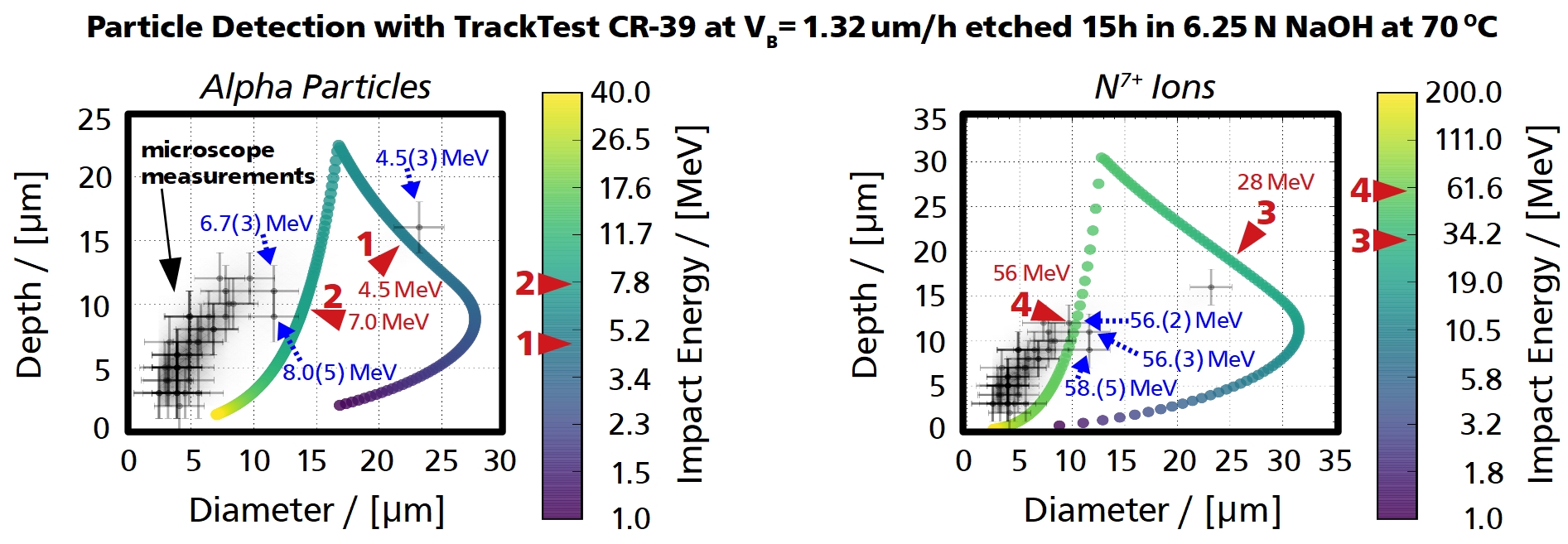}
\caption{Z-scan of etched pits on a TasTrak CR-39 after \SI{15}{\hour} of etching with $6.25\text{~N}$ NaOH etchant at \SI{70}{\celsius}. The bulk etching rate was quantified to \SI{1.32}{\micro\metre\per\hour}. Experimental data points are underlaid with a relative point number density color map in grayscale. The simulated spectrally resolving species specific $L(D)$ lobes are for alpha particles on the left hand side and for fully ionized Nitrogen ions on the right hand side. Specific impact energies are highlighted with red marker arrows (1--4). Experimental data points that overlap with the lobes in the range of their uncertainty are denoted with the closest corresponding energy on the lobe. Simulated with the CR-39 plug-in of PySTarT.}
\label{fig:CR39_zscanresults}
\end{figure*}

A second analysis strategy focuses on a Z-scan of one typical area containing all diameters of craters. We vary the focusing position of the imaging system in \si{\micro\metre} steps and follow crater walls from tip to surface to determine the length of the etched tracks. The diameter of pits is determined at the respective etched surface. Results compared to interpolations of calibration data are depicted in figure \ref{fig:CR39_zscanresults}. Data points are underlain with a relative point number density color map in gray-scale and the simulated spectrally resolving species specific $L(D)$ lobes.

% TODO ref to pystart and explain how calculus works

Most etch pits are shallow and of small diameter. Smallest diameters of \SI{3.5+-1.5}{\micro\metre} belong to craters with a large variation of depth with \SI{4.5+-3.}{\micro\metre}. Deeper tracks in a range from \SI{6.+-2.}{\micro\metre} to \SI{12.+-2.}{\micro\metre} form a linear slope with their respective diameters from \SI{4.+-2.}{\micro\metre} to \SI{10.+-2.}{\micro\metre}. Two singular data points with \SI{9.+-2.}{\micro\metre} and \SI{11.+-2.}{\micro\metre} track length do not follow this trend, both of  \SI{12.+-2.}{\micro\metre} diameter. Overall, the number density of craters decreases with increasing diameter. In the sample sequence, there is only one crater with a large diameter, here of \SI{23.+-2.}{\micro\metre} and \SI{16.+-2.}{\micro\metre} depth. From the few data points, it is difficult to conclude on a continuous relation between different types of craters.

Larger etch pits show characteristics similar to alpha particles of \SI{4.5+-0.3}{\mega\electronvolt} and smaller etch pits are in the vicinity of nitrogen projectiles of \SIrange{56+-2}{200+-50}{\mega\electronvolt}. \black{The data distribution for etch pits with diameters smaller than \SI{10}{\micro\metre} makes it difficult to distinguish wither alpha particles of \SI{6.7+-0.3}{\mega\electronvolt} to \SI{40+-14}{\mega\electronvolt} or nitrogen ions of \SIrange{56+-2}{100+-60}{\mega\electronvolt}.}

% -----------------------------------------------------------------------------------------------------------------------------------------------------------------------------

\bibliographystyle{unsrt}
\bibliography{references_2}

\begin{thebibliography}{10}

\bibitem{Strickland_1985}
Donna Strickland and Gerard Mourou.
\newblock Compression of amplified chirped optical pulses.
\newblock {\em Opt. Commun.}, 56(3):219--221, 1985.

\bibitem{Daido_2012}
Hiroyuki Daido, Mamiko Nishiuchi, and Alexander~S Pirozhkov.
\newblock Review of laser-driven ion sources and their applications.
\newblock {\em Rep. Prog. Phys.}, 75(5):056401, April 2012.

\bibitem{Macchi_2013}
Andrea Macchi, Marco Borghesi, and Matteo Passoni.
\newblock Ion acceleration by superintense laser-plasma interaction.
\newblock {\em Rev. Mod. Phys.}, 85:751--793, May 2013.

\bibitem{Borghesi_2004}
M.~Borghesi, A.~J. Mackinnon, D.~H. Campbell, D.~G. Hicks, S.~Kar, P.~K. Patel,
  D.~Price, L.~Romagnani, A.~Schiavi, and O.~Willi.
\newblock Multi-mev proton source investigations in ultraintense laser-foil
  interactions.
\newblock {\em Phys. Rev. Lett.}, 92:055003, February 2004.

\bibitem{Santos_2015}
J~J Santos, M~Bailly-Grandvaux, L~Giuffrida, P~Forestier-Colleoni, S~Fujioka,
  Z~Zhang, P~Korneev, R~Bouillaud, S~Dorard, D~Batani, M~Chevrot, J~E Cross,
  R~Crowston, J-L Dubois, J~Gazave, G~Gregori, E~d'Humi{\`{e}}res, S~Hulin,
  K~Ishihara, S~Kojima, E~Loyez, J-R Marqu{\`{e}}s, A~Morace, P~Nicolaï,
  O~Peyrusse, A~Poy{\'{e}}, D~Raffestin, J~Ribolzi, M~Roth, G~Schaumann,
  F~Serres, V~T Tikhonchuk, P~Vacar, and N~Woolsey.
\newblock Laser-driven platform for generation and characterization of strong
  quasi-static magnetic fields.
\newblock {\em New J. Phys.}, 17(8):083051, aug 2015.

\bibitem{Eh2017-1}
{M.~Ehret} et~al.
\newblock Picosecond-laser driven ultra-fast em fields propagating along coil
  targets for proton beam micro-lensing.
\newblock {\em GSI Scientific Report 2016}, GSI-2017-1:227, 2017.

\bibitem{Patel_2003}
P.~K. Patel, A.~J. Mackinnon, M.~H. Key, T.~E. Cowan, M.~E. Foord, M.~Allen,
  D.~F. Price, H.~Ruhl, P.~T. Springer, and R.~Stephens.
\newblock Isochoric heating of solid-density matter with an ultrafast proton
  beam.
\newblock {\em Phys. Rev. Lett.}, 91:125004, September 2003.

\bibitem{Roth_2001}
M.~Roth, T.~E. Cowan, M.~H. Key, S.~P. Hatchett, C.~Brown, W.~Fountain,
  J.~Johnson, D.~M. Pennington, R.~A. Snavely, S.~C. Wilks, K.~Yasuike,
  H.~Ruhl, F.~Pegoraro, S.~V. Bulanov, E.~M. Campbell, M.~D. Perry, and
  H.~Powell.
\newblock Fast ignition by intense laser-accelerated proton beams.
\newblock {\em Phys. Rev. Lett.}, 86:436--439, January 2001.

\bibitem{Ledingham_2003}
K.~W.~D. Ledingham, P.~McKenna, and R.~P. Singhal.
\newblock Applications for nuclear phenomena generated by ultra-intense lasers.
\newblock {\em Science}, 300(5622):1107--1111, 2003.

\bibitem{Spencer_2001}
I.~Spencer, K.W.D. Ledingham, R.P. Singhal, T.~McCanny, P.~McKenna, E.L. Clark,
  K.~Krushelnick, M.~Zepf, F.N. Beg, M.~Tatarakis, A.E. Dangor, P.A. Norreys,
  R.J. Clarke, R.M. Allott, and I.N. Ross.
\newblock Laser generation of proton beams for the production of short-lived
  positron emitting radioisotopes.
\newblock {\em Nucl. Instrum. Methods Phys. Res. B}, 183(3):449--458, 2001.

\bibitem{Lendingham_2014}
K.~W.~D. Ledingham et~al.
\newblock Towards laser driven hadron cancer radiotherapy: A review of
  progress.
\newblock {\em Appl. Sci.}, 4:402--443, 2014.

\bibitem{Wagner_2016}
F~Wagner, O~Deppert, C~Brabetz, P~Fiala, A~Kleinschmidt, P~Poth, VA~Schanz,
  A~Tebartz, B~Zielbauer, M~Roth, et~al.
\newblock Maximum proton energy above 85 mev from the relativistic interaction
  of laser pulses with micrometer thick ch 2 targets.
\newblock {\em Phys. Rev. Lett.}, 116(20):205002, 2016.

\bibitem{Higginson_2018}
A~Higginson, RJ~Gray, M~King, RJ~Dance, SDR Williamson, NMH Butler, R~Wilson,
  R~Capdessus, C~Armstrong, JS~Green, et~al.
\newblock Near-100 mev protons via a laser-driven transparency-enhanced hybrid
  acceleration scheme.
\newblock {\em Nat. Commun.}, 9(1):1--9, 2018.

\bibitem{Esarey_2009}
E.~{Esarey}, C.~B. {Schroeder}, and W.~P. {Leemans}.
\newblock {Physics of laser-driven plasma-based electron accelerators}.
\newblock {\em Rev. Mod. Phys.}, 81(3):1229--1285, July 2009.

\bibitem{Krushelnick_1999}
K.~{Krushelnick}, E.~L. {Clark}, Z.~{Najmudin}, M.~{Salvati}, M.~I.~K.
  {Santala}, M.~{Tatarakis}, A.~E. {Dangor}, V.~{Malka}, D.~{Neely},
  R.~{Allott}, and C.~{Danson}.
\newblock {Multi-MeV Ion Production from High-Intensity Laser Interactions with
  Underdense Plasmas}.
\newblock {\em Phys. Rev. Lett.}, 83(4):737--740, July 1999.

\bibitem{Willingale_2006}
L.~Willingale, S.~P.~D. Mangles, P.~M. Nilson, R.~J. Clarke, A.~E. Dangor,
  M.~C. Kaluza, S.~Karsch, K.~L. Lancaster, W.~B. Mori, Z.~Najmudin,
  J.~Schreiber, A.~G.~R. Thomas, M.~S. Wei, and K.~Krushelnick.
\newblock Collimated multi-mev ion beams from high-intensity laser interactions
  with underdense plasma.
\newblock {\em Phys. Rev. Lett.}, 96:245002, June 2006.

\bibitem{Lifschitz_2014}
A~Lifschitz, F~Sylla, S~Kahaly, A~Flacco, M~Veltcheva, G~Sanchez-Arriaga,
  E~Lefebvre, and V~Malka.
\newblock Ion acceleration in underdense plasmas by ultra-short laser pulses.
\newblock {\em New J. Phys.}, 16(3):033031, March 2014.

\bibitem{Willingale_2009}
L.~Willingale, S.~R. Nagel, A.~G.~R. Thomas, C.~Bellei, R.~J. Clarke, A.~E.
  Dangor, R.~Heathcote, M.~C. Kaluza, C.~Kamperidis, S.~Kneip, K.~Krushelnick,
  N.~Lopes, S.~P.~D. Mangles, W.~Nazarov, P.~M. Nilson, and Z.~Najmudin.
\newblock Characterization of high-intensity laser propagation in the
  relativistic transparent regime through measurements of energetic proton
  beams.
\newblock {\em Phys. Rev. Lett.}, 102:125002, March 2009.

\bibitem{Fiuza_2012}
{Fi{\'u}za, Frederico and Stockem, Anne and Boella, Elisabetta and Fonseca, RA
  and Silva, LO and Haberberger, D and Tochitsky, Sergei and Gong, Chao and
  Mori, Warren B and Joshi, Chandrasekar}.
\newblock Laser-driven shock acceleration of monoenergetic ion beams.
\newblock {\em Phys. Rev. Lett.}, 109(21):215001, 2012.

\bibitem{Debayle_2017}
A~Debayle, F~Mollica, B~Vauzour, Y~Wan, A~Flacco, V~Malka, X~Davoine, and
  L~Gremillet.
\newblock Electron heating by intense short-pulse lasers propagating through
  near-critical plasmas.
\newblock {\em New J. Phys.}, 19(12):123013, December 2017.

\bibitem{Rosmej_2019}
O~N Rosmej, N~E Andreev, S~Zaehter, N~Zahn, P~Christ, B~Borm, T~Radon,
  A~Sokolov, L~P Pugachev, D~Khaghani, F~Horst, N~G Borisenko, G~Sklizkov, and
  V~G Pimenov.
\newblock Interaction of relativistically intense laser pulses with long-scale
  near critical plasmas for optimization of laser based sources of {MeV}
  electrons and gamma-rays.
\newblock {\em New J. Phys.}, 21(4):043044, April 2019.

\bibitem{Pazzaglia_2020}
Andrea {Pazzaglia}, Luca {Fedeli}, Arianna {Formenti}, Alessandro {Maffini},
  and Matteo {Passoni}.
\newblock {A theoretical model of laser-driven ion acceleration from
  near-critical double-layer targets}.
\newblock {\em Commun. Phys.}, 3(1):133, August 2020.

\bibitem{Moiseev_1963}
S.~S. {Moiseev} and R.~Z. {Sagdeev}.
\newblock {Collisionless shock waves in a plasma in a weak magnetic field}.
\newblock {\em J. Nucl. Energy}, 5(1):43--47, January 1963.

\bibitem{Forslund_1970}
D.~W. {Forslund} and C.~R. {Shonk}.
\newblock {Formation and Structure of Electrostatic Collisionless Shocks}.
\newblock {\em Phys. Rev. Lett.}, 25(25):1699--1702, December 1970.

\bibitem{Sorasio_2006}
G~Sorasio, Michael Marti, Ricardo Fonseca, and Luis~O Silva.
\newblock Very high mach-number electrostatic shocks in collisionless plasmas.
\newblock {\em Phys. Rev. Lett.}, 96(4):045005, 2006.

\bibitem{Cairns_2014}
R.~A. {Cairns}, R.~{Bingham}, P.~{Norreys}, and R.~{Trines}.
\newblock {Laminar shocks in high power laser plasma interactions}.
\newblock {\em Phys. Plasmas}, 21(2):022112, February 2014.

\bibitem{Silva_2004}
Lu\'{\i}s~O. Silva, Michael Marti, Jonathan~R. Davies, Ricardo~A. Fonseca,
  Chuang Ren, Frank~S. Tsung, and Warren~B. Mori.
\newblock Proton shock acceleration in laser-plasma interactions.
\newblock {\em Phys. Rev. Lett.}, 92:015002, January 2004.

\bibitem{Fiuza_2013}
F.~{Fiuza}, A.~{Stockem}, E.~{Boella}, R.~A. {Fonseca}, L.~O. {Silva},
  D.~{Haberberger}, S.~{Tochitsky}, W.~B. {Mori}, and C.~{Joshi}.
\newblock {Ion acceleration from laser-driven electrostatic shocksa)}.
\newblock {\em Phys. Plasmas}, 20(5):056304, May 2013.

\bibitem{Kim_2016}
Young-Kuk Kim, Teyoun Kang, Moon~Youn Jung, and Min~Sup Hur.
\newblock Effects of laser polarizations on shock generation and shock ion
  acceleration in overdense plasmas.
\newblock {\em Phys. Rev. E}, 94(3):033211, 2016.

\bibitem{Liu_2016}
M.~Liu, S.~M. Weng, Y.~T. Li, D.~W. Yuan, M.~Chen, P.~Mulser, Z.~M. Sheng,
  M.~Murakami, L.~L. Yu, X.~L. Zheng, and J.~Zhang.
\newblock Collisionless electrostatic shock formation and ion acceleration in
  intense laser interactions with near critical density plasmas.
\newblock {\em Phys. Plasmas}, 23(11):113103, 2016.

\bibitem{Pak_2018}
A.~Pak, S.~Kerr, N.~Lemos, A.~Link, P.~Patel, F.~Albert, L.~Divol, B.~B.
  Pollock, D.~Haberberger, D.~Froula, M.~Gauthier, S.~H. Glenzer, A.~Longman,
  L.~Manzoor, R.~Fedosejevs, S.~Tochitsky, C.~Joshi, and F.~Fiuza.
\newblock Collisionless shock acceleration of narrow energy spread ion beams
  from mixed species plasmas using $\mu$m lasers.
\newblock {\em Phys. Rev. Accel. Beams}, 21:103401, October 2018.

\bibitem{Antici_2017}
P.~{Antici}, E.~{Boella}, S.~N. {Chen}, D.~S. {Andrews}, M.~{Barberio},
  J.~{B{\"o}ker}, F.~{Cardelli}, J.~L. {Feugeas}, M.~{Glesser},
  P.~{Nicola{\"\i}}, L.~{Romagnani}, M.~{Scisci{\`o}}, M.~{Starodubtsev},
  O.~{Willi}, J.~C. {Kieffer}, V.~{Tikhonchuk}, H.~{P{\'e}pin}, L.~O. {Silva},
  E.~d' {Humi{\`e}res}, and J.~{Fuchs}.
\newblock {Acceleration of collimated 45 MeV protons by collisionless shocks
  driven in low-density, large-scale gradient plasmas by a $10^{20}\,W/cm^2$,
  $1\,\rm \mu m$ laser}.
\newblock {\em Sci. Rep.}, 7:16463, November 2017.

\bibitem{Chen_2017}
S.~N. {Chen}, M.~{Vranic}, T.~{Gangolf}, E.~{Boella}, P.~{Antici},
  M.~{Bailly-Grandvaux}, P.~{Loiseau}, H.~{P{\'e}pin}, G.~{Revet}, J.~J.
  {Santos}, A.~M. {Schroer}, Mikhail {Starodubtsev}, O.~{Willi}, L.~O. {Silva},
  E.~{d'Humi{\`e}res}, and J.~{Fuchs}.
\newblock {Collimated protons accelerated from an overdense gas jet irradiated
  by a $1\,\rm \mu m$ wavelength high-intensity short-pulse laser}.
\newblock {\em Sci. Rep.}, 7:13505, October 2017.

\bibitem{Puyuelo_2019}
P.~Puyuelo-Valdes, J.~L. Henares, F.~Hannachi, T.~Ceccotti, J.~Domange,
  M.~Ehret, E.~d'Humieres, L.~Lancia, J.-R. Marqu{\`{e}}s, X.~Ribeyre, J.~J.
  Santos, V.~Tikhonchuk, and M.~Tarisien.
\newblock Proton acceleration by collisionless shocks using a supersonic h2
  gas-jet target and high-power infrared laser pulses.
\newblock {\em Phys. Plasmas}, 26(12):123109, 2019.

\bibitem{Bulanov_2010}
Stepan~S. Bulanov, Valery~Yu. Bychenkov, Vladimir Chvykov, Galina Kalinchenko,
  Dale~William Litzenberg, Takeshi Matsuoka, Alexander G.~R. Thomas, Louise
  Willingale, Victor Yanovsky, Karl Krushelnick, and Anatoly Maksimchuk.
\newblock Generation of gev protons from 1 pw laser interaction with near
  critical density targets.
\newblock {\em Phys. Plasmas}, 17(4):043105, 2010.

\bibitem{Nakamura_2010}
Tatsufumi Nakamura, Sergei~V. Bulanov, Timur~Zh. Esirkepov, and Masaki Kando.
\newblock High-energy ions from near-critical density plasmas via magnetic
  vortex acceleration.
\newblock {\em Phys. Rev. Lett.}, 105:135002, September 2010.

\bibitem{Sylla_2012_2}
F.~Sylla, A.~Flacco, S.~Kahaly, M.~Veltcheva, A.~Lifschitz, G.~Sanchez-Arriaga,
  E.~Lefebvre, and V.~Malka.
\newblock Anticorrelation between ion acceleration and nonlinear coherent
  structures from laser-underdense plasma interaction.
\newblock {\em Phys. Rev. Lett.}, 108:115003, March 2012.

\bibitem{Helle_2016}
M.~H. Helle, D.~F. Gordon, D.~Kaganovich, Y.~Chen, J.~P. Palastro, and A.~Ting.
\newblock Laser-accelerated ions from a shock-compressed gas foil.
\newblock {\em Phys. Rev. Lett.}, 117:165001, October 2016.

\bibitem{Park_2019}
J.~Park, S.~S. Bulanov, J.~Bin, Q.~Ji, S.~Steinke, J.-L. Vay, C.~G.~R. Geddes,
  C.~B. Schroeder, W.~P. Leemans, T.~Schenkel, and E.~Esarey.
\newblock Ion acceleration in laser generated megatesla magnetic vortex.
\newblock {\em Phys. Plasmas}, 26(10):103108, 2019.

\bibitem{dHumieres_2013_1}
{E d'Humi\`{e}res and P Antici and M Glesser and J Boeker and F Cardelli and S
  Chen and J L Feugeas and F Filippi and M Gauthier and A Levy and P
  Nicola\"{i} and H P\'{e}pin and L Romagnani and M Scisci\`{o} and V T
  Tikhonchuk and O Willi and J C Kieffer and J Fuchs}.
\newblock Investigation of laser ion acceleration in low-density targets using
  exploded foils.
\newblock {\em Plasma Phys. Control. Fusion}, 55(12):124025, November 2013.

\bibitem{Gauthier_2014}
{Gauthier,M. and L\'{e}vy,A. and d'Humi\`{e}res,E. and Glesser,M. and
  Albertazzi,B. and Beaucourt,C. and Breil,J. and Chen,S. N. and Dervieux,V.
  and Feugeas,J. L. and Nicola\"{i},P. and Tikhonchuk,V. and P\'{e}pin,H. and
  Antici,P. and Fuchs,J.}
\newblock Investigation of longitudinal proton acceleration in exploded targets
  irradiated by intense short-pulse laser.
\newblock {\em Phys. Plasmas}, 21(1):013102, 2014.

\bibitem{Prencipe_2016}
{I Prencipe and A Sgattoni and D Dellasega and L Fedeli and L Cialfi and Il Woo
  Choi and I Jong Kim and K A Janulewicz and K F Kakolee and Hwang Woon Lee and
  Jae Hee Sung and Seong Ku Lee and Chang Hee Nam and M Passoni}.
\newblock Development of foam-based layered targets for laser-driven ion beam
  production.
\newblock {\em Plasma Phys. Controlled Fusion}, 58(3):034019, February 2016.

\bibitem{Bin_2015}
J.~H. {Bin}, W.~J. {Ma}, H.~Y. {Wang}, M.~J.~V. {Streeter}, C.~{Kreuzer},
  D.~{Kiefer}, M.~{Yeung}, S.~{Cousens}, P.~S. {Foster}, B.~{Dromey}, X.~Q.
  {Yan}, R.~{Ramis}, J.~{Meyer-ter-Vehn}, M.~{Zepf}, and J.~{Schreiber}.
\newblock {Ion Acceleration Using Relativistic Pulse Shaping in
  Near-Critical-Density Plasmas}.
\newblock {\em Phys. Rev. Lett.}, 115(6):064801, August 2015.

\bibitem{Bin_2018}
J.~H. {Bin}, M.~{Yeung}, Z.~{Gong}, H.~Y. {Wang}, C.~{Kreuzer}, M.~L. {Zhou},
  M.~J.~V. {Streeter}, P.~S. {Foster}, S.~{Cousens}, B.~{Dromey},
  J.~{Meyer-ter-Vehn}, M.~{Zepf}, and J.~{Schreiber}.
\newblock {Enhanced Laser-Driven Ion Acceleration by Superponderomotive
  Electrons Generated from Near-Critical-Density Plasma}.
\newblock {\em Phys. Rev. Lett.}, 120(7):074801, February 2018.

\bibitem{Goers_2015}
A.~J. Goers, G.~A. Hine, L.~Feder, B.~Miao, F.~Salehi, J.~K. Wahlstrand, and
  H.~M. Milchberg.
\newblock Multi-mev electron acceleration by subterawatt laser pulses.
\newblock {\em Phys. Rev. Lett.}, 115:194802, November 2015.

\bibitem{Henares_2019}
J.~L. Henares, P.~Puyuelo-Valdes, F.~Hannachi, T.~Ceccotti, M.~Ehret, F.~Gobet,
  L.~Lancia, J.-R. Marqu{\`{e}}s, J.~J. Santos, M.~Versteegen, and M.~Tarisien.
\newblock Development of gas jet targets for laser-plasma experiments at
  near-critical density.
\newblock {\em Rev. Sci. Instrum.}, 90(6):063302, 2019.

\bibitem{Sylla_2012_1}
F~Sylla, M~Veltcheva, S~Kahaly, Alessandro Flacco, and Victor Malka.
\newblock Development and characterization of very dense submillimetric gas
  jets for laser-plasma interaction.
\newblock {\em Rev. Sci. Instrum.}, 83(3):033507, 2012.

\bibitem{Volpe_2019}
L.~Volpe, R.~Fedosejevs, G.~Gatti, J.~A. P{\'{e}}rez-Hern{\'{a}}ndez,
  C.~M{\'{e}}ndez, J.~Api{\~{n}}aniz, X.~Vaisseau, C.~Salgado, M.~Huault,
  S.~Malko, and et~al.
\newblock Generation of high energy laser-driven electron and proton sources
  with the 200 tw system vega 2 at the centro de laseres pulsados.
\newblock {\em High Power Laser Sci. Eng.}, 7:e25, 2019.

\bibitem{Chanteloup_2005}
J.~Primot et~al.
\newblock L'analyse de surface d'onde par interf{\'{e}}rom{\'{e}}trie {\`{a}}
  d{\'{e}}calage multilat{\'{e}}ral.
\newblock {\em Photoniques (Orsay)}, 2005.

\bibitem{Hutchinson_2002}
I.~H. Hutchinson.
\newblock {\em Principles of Plasma Diagnostics}.
\newblock Cambridge University Press, 2 edition, 2002.

\bibitem{Nue2009}
F.~N\"{u}rnberg et~al.
\newblock Radiochromic film imaging spectroscopy of laser-accelerated proton
  beams.
\newblock {\em Rev. Sci. Instrum.}, 80(033301), March 2009.

\bibitem{Yo1958}
D.~A. Young.
\newblock Etching of radiation damage in lithium fluoride.
\newblock {\em Nature}, 182(4632):375--377, August 1958.

\bibitem{DB1987}
S.~A. Durrani and R.~K. Bull.
\newblock {\em Solid State Nuclear Track Detection}.
\newblock Pergamon, 1987.

\bibitem{Ha2008}
M.~Hajek et~al.
\newblock Convolution of tld and ssntd measurements during the brados-1
  experiment onboard iss (2001).
\newblock {\em Radiat. Meas.}, 43(7):1231--1236, 2008.

\bibitem{GEBT3}
L.~Porter.
\newblock {\em GAFCHROMIC DOSIMETRY MEDIA, TYPE EBT-3}.
\newblock Ashland, February 2016.

\bibitem{Note1}
Films were made available to us by the Advanced Materials Group of Ashland
  Specialty Ingredients G.P., 1005 US Hwy No 202/206, Bridgewater, NJ 08807,
  USA.

\bibitem{Me1996}
K.~Mehta and I.~Janovský.
\newblock Measurements of electron depth-dose distributions in thick plastics
  and effects of accumulated charge.
\newblock {\em Radiat. Phys. Chem.}, 47(3):487 -- 490, 1996.
\newblock Tihany Symposium on Radiation Chemistry.

\bibitem{Lefebvre2003_NucFus}
E~Lefebvre, N~Cochet, S~Fritzler, Victor Malka, M-M Al{\'e}onard, J-F Chemin,
  S~Darbon, L~Disdier, J{\'e}r{\^o}me Faure, A~Fedotoff, et~al.
\newblock Electron and photon production from relativistic laser--plasma
  interactions.
\newblock {\em Nucl. Fusion}, 43(7):629, 2003.

\bibitem{Jo2019}
L.~Jonu\v{s}auskas et~al.
\newblock Mesoscale laser 3d printing.
\newblock {\em Opt. Express}, 27(11):15205--15221, May 2019.

\bibitem{Saini1987}
R.D. Saini and P.K. Bhattacharyya.
\newblock Radiolytic oxidation of u(iv) sulphate in aqueous solution by alpha
  particles from cyclotron.
\newblock {\em Int. J. Radiat. Appl. Instrum. C.}, 29(5):375 -- 379, 1987.

\bibitem{Ba2008}
G.~Baldacchino.
\newblock Pulse radiolysis in water with heavy-ion beams. a short review.
\newblock {\em Radiat. Phys. Chem.}, 77(10):1218 -- 1223, 2008.
\newblock The International Symposium on Charged Particle and Photon
  Interaction with Matter - ASR 2007.

\bibitem{Bo2020}
P.~Boller, A.~Zylstra, P.~Neumaye, et~al.
\newblock First on-line detection of radioactive fission isotopes produced by
  laser-accelerated protons.
\newblock {\em Sci. Rep.}, (10):17183, 2020.

\bibitem{Ch2018}
J.~Chatal et~al.
\newblock Alphatherapy, the new impetus to targeted radionuclide therapy?
\newblock {\em Eur. J. Nucl. Med. Mol. Imaging}, (45):1362--1363, 2018.

\bibitem{IAEA_HHR_15}
IAEA.
\newblock Medical physics staffing needs in diagnostic imaging and radionuclide
  therapy: an activity based approach.
\newblock Technical report, 2018.

\bibitem{Ma2018}
M.~Makvandi et~al.
\newblock Alpha-emitters and targeted alpha therapy in oncology: from basic
  science to clinical investigations.
\newblock {\em Target. Oncol.}, (13):189--203, 2018.

\bibitem{corallo1980x}
DM~Corallo, DM~Creek, and GM~Murray.
\newblock The x-ray calibration of silicon pin diodes between 1.5 and 17.4 kev.
\newblock {\em J. Phys. E: Sci. Instrum.}, 13(6):623, 1980.

\bibitem{Zi1999}
J.~F. Ziegler.
\newblock The stopping of energetic light ions in elemental matter.
\newblock {\em Rev. Appl. Phys.}, 85:1249--1272, 1999.

\bibitem{Si2006}
P.~Sigmund.
\newblock {\em Particle Penetration and Radiation Effects}.
\newblock Number 151 in Solid State Sciences. Berlin Heidelberg :
  Springer-Verlag, 2006.

\bibitem{Ma2015}
J.~A.~Mart{\'{i}}n-Viera Cueto et~al.
\newblock A universal dose--response curve for radiochromic films.
\newblock {\em Med. Phys.}, 42(1):221--231, 2015.

\bibitem{SP1986}
J.~Stejny and T.~Portwood.
\newblock A study of the molecular structure in cr-39.
\newblock {\em Int. J. Radiat. Appl. Instrum. Part D}, 12(1):121--123, 1986.
\newblock Special Volume Solid State Nuclear Track Detectors.

\bibitem{Ma1987}
W.~Maurer.
\newblock Neutron and gamma irradiation effects onorganic insulating materials
  for fusionmagnets.
\newblock TECDOC 417, KIT, International Atomic Energy Agency, Vienna, 1987.

\bibitem{Th1995}
N.~X. Thang and T.~T. Doan.
\newblock Structure effect in response function of cr-39 detector.
\newblock {\em Radiat. Meas.}, 25(1):185--187, 1995.
\newblock Nuclear Tracks in Solids.

\bibitem{Ch1997}
C.~S. Chong et~al.
\newblock Uv-vis and ftir spectral studies of cr-39 plastics irradiated with
  x-rays.
\newblock {\em Radiat. Meas.}, 28(1):119--122, 1997.
\newblock International Conference on Nuclear Tracks in Solids.

\bibitem{Ch1988}
J.~Charv\'{a}t and F.~Spurn\'{y}.
\newblock Etching characteristics of cellulose nitrate and cr-39 after high
  dose electron irradiation.
\newblock {\em Int. J. Radiat. Appl. Instrum. Part D}, 14(4):451--455, 1988.

\bibitem{TASTRAKMSDS}
A.~Worley.
\newblock Tastrak material safety data sheet msds-070314-1 (revision 12).
\newblock Technical report, TASL, Napier House Meadow Grove Shirehampton
  Bristol BS11 9PJ, 2014.

\bibitem{Zh2019}
Y.~Zhang et~al.
\newblock Energy calibration of a cr-39 nuclear-track detector irradiated by
  charged particles.
\newblock {\em Nucl. Sci. Tech.}, (30):87, 2019.

\bibitem{Az2012}
A.~A. Azooz, S.~H. Al-Nia\'{}emi, and M.~A. Al-Jubbori.
\newblock Empirical parameterization of cr-39 longitudinal track depth.
\newblock {\em Radiat. Meas.}, 47(1):67--72, 2012.

\bibitem{Az2013}
A.~A. Azooz, D.~Hermsdorf, and M.~A. Al-Jubbori.
\newblock New approach of modeling charged particles track development in cr-39
  detectors.
\newblock {\em Radiat. Meas.}, 58:94--100, 2013.

\end{thebibliography}

\end{document}